\documentclass[twocolumn,floatfix,showpacs,preprintnumbers,amsmath]{revtex4}

\usepackage{amssymb,graphics}
\usepackage{color}
\usepackage{psfrag}
\usepackage{graphicx}

\begin{document}

\title{Non-linear oscillatory rheological properties of a generic continuum foam model: comparison with experiments and shear-banding predictions}
\author{Sylvain B\'ENITO}
\affiliation{Universit\'e Bordeaux~1, INRIA Futurs projet MC2 et IMB, 351 Cours de la Lib\'eration, F--33405 TALENCE cedex, France}
\author{Fran\c{c}ois MOLINO}
\affiliation{Institut de G\'enomique Fonctionnelle, Department of Endocrinology, CNRS, UMR~5203, INSERM U661, Universit\'e Montpellier Sud de France, 141 Rue de la Cardonille, F--34094 MONTPELLIER cedex 05, France }
\affiliation{Academy of Bradylogists}
\author{Charles-Henri BRUNEAU}
\author{Thierry COLIN}
\affiliation{Universit\'e Bordeaux~1, INRIA Futurs projet MC2 et IMB, 351 Cours de la Liberation, F--33405 TALENCE cedex, France}
\author{Cyprien GAY}
\affiliation{Mati\`ere et Syst\`emes Complexes (MSC), CNRS UMR 7057, Universit\'e Paris Diderot--Paris~7, B\^atiment Condorcet, Case courrier 7056, 75205 Paris Cedex 13}
\affiliation{Academy of Bradylogists}                     
%
%

\date{\today}
\begin{abstract}
The occurence of shear bands in a complex fluid is generally understood 
as resulting from a structural evolution of the material under shear, 
which leads (from a theoretical perspective) to a non-monotonic stationnary flow curve 
related to the coexistence of different states of the material under shear. 
In this paper we present a scenario for shear-banding in a particular class 
of complex fluids, namely foams and concentrated emulsions, 
which differs from other scenarii in two important ways.
First, the appearance of shear bands is shown to be possible 
both without any intrinsic physical evolution of the material 
(e.g. {\em via} a parameter coupled to the flow such as concentration or entanglements)
and without any finite critical shear rate 
below which the flow does not remain stationary and homogeneous.
Secondly, the appearance of shear bands depends on the initial conditions, 
{\em i.e.}, the preparation of the material. 
In other words, it is history dependent.
This behaviour relies on the tensorial character of the underlying model (2D or 3D)
and is triggered by an initially inhomogeneous strain distribution in the material.
The shear rate displays a discontinuity at the band boundary,
whose amplitude is history dependent
and thus depends on the sample preparation.
\end{abstract}


\maketitle

\newcommand{\hs}{\hspace{0.7cm}}
\newcommand{\be}{\begin{equation}}
\newcommand{\ee}{\end{equation}}
\newcommand{\bee}{\begin{eqnarray}}
\newcommand{\eee}{\end{eqnarray}}
\newcommand{\fin}{\nonumber\\}
\newcommand{\upperconv}[1]{^\nabla{#1}}
\newcommand{\lowerconv}[1]{\frac{{\rm D^{(-)}}{#1}}{{\rm D}t}}
\newcommand{\trace}{{\rm tr}}
\newcommand{\transp}[1]{{#1}^{\rm T}}
\newcommand{\transpinv}[1]{{#1}^{\rm -T}}
\newcommand{\inv}[1]{{#1}^{-1}}
\newcommand{\letout}[1]{\widehat{#1}}
\newcommand{\rhozero}{\rho_0}
\newcommand{\rhoun}{\rho^1}
\newcommand{\rhounzero}{\rho^1_0}
\newcommand{\rhodeux}{\rho^2}
\newcommand{\rhodeuxzero}{\rho^2_0}
\newcommand{\rholiq}{\rho_{\rm liq}}
\newcommand{\phiun}{\phi_1}
\newcommand{\phideux}{\phi_2}
\newcommand{\etaun}{\eta_1}
\newcommand{\etadeux}{\eta_2}
\newcommand{\jun}{{\vec{j}}^1}
\newcommand{\jdeux}{{\vec{j}}^2}
\newcommand{\patm}{p_{\rm atm}}
\newcommand{\pun}{p_1}
\newcommand{\pdeux}{p_2}
\newcommand{\permeation}{\kappa}
\newcommand{\permeationun}{\permeation_1}
\newcommand{\permeationdeux}{\permeation_2}
\newcommand{\soif}{[\rm thirst]}
\newcommand{\unittensor}{{\rm I}}
\newcommand{\G}{G}
\newcommand{\W}{E}
\newcommand{\WW}{W}
\newcommand{\mra}{k_1}
\newcommand{\mrb}{k_2}
\newcommand{\energy}{U}
\newcommand{\pdissip}{P_{\rm dissip}}
\newcommand{\screated}{S_{\rm created}}
\newcommand{\moduletensoriel}{{\cal G}}
\newcommand{\stress}{{\cal G}}
\newcommand{\siy}{{\sigma_y}}
\newcommand{\Sis}{{\siy}}
\newcommand{\compl}{{\cal C}}
\newcommand{\defse}{{\cal E}}
\newcommand{\se}{{\epsilon_{\rm s}}}
\newcommand{\dse}{{\rm d}\se}
\newcommand{\Dse}{\dot{\se}}
\newcommand{\vX}{\vec{X}}
\newcommand{\vx}{\vec{x}}
\newcommand{\FF}{F}
\newcommand{\Cauchy}{C}
\newcommand{\LCauchy}{C^-} 
\newcommand{\eLCauchy}{e^-} 
\newcommand{\Finger}{{B}}  
\newcommand{\eFinger}{{e}}
\newcommand{\vxp}{{\vec{x}^\prime}}
\newcommand{\vxpt}{{\vec{x}^{\prime\,{\rm T}}}}
\newcommand{\vv}{\vec{v}}
\newcommand{\gradv}{\nabla\vv}
\newcommand{\si}{\sigma}
\newcommand{\siel}{\si_{\rm el}}
\newcommand{\sid}{{\si_d}}
\newcommand{\s}{\bar{\sigma}}
\newcommand{\suu}{\s_{11}}
\newcommand{\sud}{\s_{12}}
\newcommand{\sdd}{\s_{22}}
\newcommand{\su}{\s_{1}}
\newcommand{\sd}{\s_{2}}
\newcommand{\strois}{\s_{3}}
\newcommand{\dsi}{{\rm d}\si}
\newcommand{\sip}{\dot{\si}}
\newcommand{\Dsi}{\sip}
\newcommand{\sipl}{\si_{\rm pl}}
\newcommand{\sippl}{\sip_{\rm pl}}
\newcommand{\dt}{{\rm d}t}
\newcommand{\uu}{u}
\newcommand{\tu}{\transp{\uu}}
\newcommand{\U}{U}
\newcommand{\T}{T}
\newcommand{\gd}{\dot{\gamma}}
\newcommand{\gdc}{{\gd_c}}
\newcommand{\est}{{e_{\rm e}}}
\newcommand{\eps}{\varepsilon}
\newcommand{\epsp}{D}
\newcommand{\epspe}{\frac{{\rm D}^+\eFinger}{{\rm D}t}}
\newcommand{\epspp}{{D_{\rm p}^B}}
\newcommand{\epsppt}{\tilde{\epspp}}
\newcommand{\omp}{\dot{\omega}}
\newcommand{\dev}{{\rm dev}}
\newcommand{\rate}{\Gamma}
\newcommand{\ct}{c}
\newcommand{\st}{s}
\newcommand{\hide}[1]{}
\newcommand{\amplepspp}{{\cal A}}
\newcommand{\weiss}{{\rm We}}


\section{Shear bands and foam rheology}

\subsection{Shear bands in complex fluids}
\label{SubSec:shb_complex_fluids}

It may seem paradoxical that a single material,
when submitted to a uniform shear stress $\sigma_{xy}$, 
between two parallel plates or two coaxial cylinders, 
may be observed simultaneously in two distinct states 
in different regions of the flow. 
This 'shear bands' observation has nevertheless become common
since the early 1990s in a variety of complex fluids: 
they appear and are 
stable~\cite{banding_flowcurve,banding_birefringence,banding_phasetransition,banding_jamming_review}, 
or sometimes fluctuate~\cite{salmon_2003_051503,salmon_2003_051504,becu,banding_heterogeneous}. 
These bands are most of the time parallel with the shearing plates~\cite{banding_flowcurve},
with a different shear rate in each band.

The current understanding of these observations 
relies in general on two essential ingredients~: 
{\em (i)} a structural evolution of the material under shear, 
and {\em (ii)} a stress response that decreases as a function of the shear rate (within a particular range).
This decrease is the mechanical signature of the structural evolution of the fluid
and is the source of the mechanical instability
that triggers the appearance of bands~\cite{lerouge_berret_2010_AdvPolymSci}.

In polymer melts or entangled polymer solutions~\cite{ShiQingWang_PRL_96_196001} 
and in entangled giant micelle solutions~\cite{lerouge_berret_2010_AdvPolymSci}, 
the flow elongates the objects,
which alters the apparent viscosity of the material
(which must be evaluated after subtracting the effect of 
wall slip~\cite{Hayes_PRL_2008_101_218301}). 
The fact that this viscosity {\em goes down} 
is principally due to the average orientation
of the objects in the shear flow.

In lyotropic lamellar phases, the transition can be associated 
with the reorganisation of the films into onion-like multilamellar 
vesicle systems~\cite{diat_1993_1427,diat_1995_3296,salmon_2003_051503,salmon_2003_051504},
also exhibiting wall slip behavior~\cite{salmon_2003_051503}.
In micellar cubic crystals the transition consists 
in an ordering of the initial polycrystal 
into a single crystal with specific planes 
becoming aligned with the plates~\cite{molino00_201,molino00_6759}.
In the last two cases, no microscopic interpretation 
of the decrease in effective viscosity occuring during the transition is currently available.

In granular materials, surface flow is a particular case of shear bands: 
the lower band is in this case blocked (zero shear)
while the flowing region is sheared. 
Again, no complete structural description is available. 
Nevertheless, it is admitted that through {\em dilatancy},
which reflects the necessity for the grains to move a little bit apart 
in order to move past each other~\cite{reynolds_centenary_1985_469}, 
the shear generates a difference in volume fraction 
between the flowing region and the blocked one.
This lower volume fraction tends 
to facilitate the flow in the flowing region even more 
as compared to the blocked region, thus stabilizing shear-banding.
When it is present, gravity is of course essential: 
it favours this phenomenon by helping the system 
segregate into a dense, blocked region
(located at the bottom if the particles are denser than the fluid)
and a less dense, flowing region.
Thus, it allows to determine the concentration profile~\cite{lenoble:073303}.

In foams and emulsions, the situation is less clear.
Shear bands were observed in 2D~\cite{debregeas_tabuteau_di_meglio,rodts_2005_69_636}.
In some cases, the observed shear-banding 
could result trivially from shear stress inhomogeneity,
due to cylindical Couette geometry ($\sigma(r)\propto 1/r^2$)
or enhanced (for 2D foams) by the presence of at least one solid 
boundary~\cite{Janiaud-hutzler-weaire-PRL2006, wang_PRE_2006_73_031401, langlois-hutzler-weaire-PRE2008-021401, katgert_mobius_vanhecke_PRL2008_disorder}
whose friction on the foam implies that $r^2\sigma(r)$ is not uniform.
In some more interesting cases, the shear rate is spatially discontinuous
at the boundary between the blocked and the sheared
regions~\cite{lauridsen_PRL2004_93_018303,rodts_2005_69_636}.
This discontinuity may arise from an apparently intrinsic impossibility for the foam
to be deformed homogeneously at low shear rates~\cite{dacruz_PRE2002_66_051305},
which then implies the presence of shear bands at low shear rates.
But at least in some cases, the shear rate at the boundary
is not unique for a given system~\cite{lauridsen_PRL2004_93_018303}
and is thus not intrinsic.
Recently, the very existence of such a finite shear rate at the boundary
has been seriously questioned~\cite{ovarlez_EPL2010_91_68005}.
As we shall see, the present work highlights yet another (history dependent) possible origin
of the shear rate spatial discontinuity,
arising from the tensorial character of the material response.
%
For these materials also, no complete structural description
accounts for flow localization in a satisfactory manner.
Dilatancy, which corresponds to a local change in water concentration $\phi$,
certainly plays an important role by easing the relative motion of bubbles or drops,
although it behaves somewhat differently from granular materials depending 
on the liquid fraction~\cite{weaire_2003_2747,rioual_2005_117,dilatancy_geometry_letter_2008}.
The structural disorder is also invoked
as a parameter coupled to the flow~\cite{kabla_scheibert_debregeas_2007b}. 
In both cases, the local fluidity
(ratio of the shear rate and the shear stress) is enhanced.

\subsection{Foam rheology}
\label{Sec:foam_rheo}

The specificity of foams as compared to other materials
is the following: not only do they flow substantially only above some threshold stress,
but they also undergo large elastic deformations at lower stress.
Hence, classical models such as visco-elastic fluids (well suited for polymeric fluids)
and elasto-plastic solids (well suited for metals)
are unsufficient to capture the behaviour of foams and emulsions.
In the past few years, much effort has been devoted
to address this challenge and describe the richer mechanical behaviour of foams.
Several rheological models have thus emerged~\cite{labiausse_2007_479,saramito,marmottant_2008_EPJEII,benito_2008_225_elasto_visco_plastic_foam,raufaste_2010_0732}.
They all assume that the foam is essentially incompressible.
Within this perimeter, some models are purely visco-elastic 
yet with a non-linear elasticity~\cite{labiausse_2007_479}.
As for the models incorporating plasticity,
they can be dispatched into two categories:
the creep term either depends on the stress 
and deformation rate~\cite{marmottant_graner,marmottant_2008_EPJEII,raufaste_2010_0732}
or on the stress only~\cite{saramito,benito_2008_225_elasto_visco_plastic_foam}.
Finally, these models also differ in the tensorial form of elasticity and creep,
a feature which is relevant for non strictly 2D systems (or for compressible materials).

Despite this variety of models, most current experiments
are not sufficiently stringent to fully test these models
and decide which ingredients are indeed relevant 
to describe the mechanical response of foams.
For instance, classical linear rheology experiments,
particularly oscillatory measurements,
are clearly unable to provide much information
about the behaviour under large stress.

Because the constitutive objects of foams are macroscopic
and can be observed directly,
statistical tools have been elaborated to measure the local deformation and deformation rate.
Using these tools, more complex geometries such as flows around obstacles
are also used in order to subject the foam
to a tensorially broader variety of sollicitations~\cite{graner_2008_369}.
Yet because these experiments are conducted in a confined geometry,
the unknown friction with the walls
and the fact that no direct measurement of the total stress is conducted,
make it difficult to test stress predictions beyond low velocities.

To complement this, in order to test the full time response of the models
even within classical geometries such as those available in a rheometer,
a broad range of experiments could be elaborated 
by choosing many different forms for the time dependence
of the applied deformation or stress.
As a first step towards this goal, 
considering the full time-dependent response
of a foam subjected to large amplitude oscillatory shear
(not only the usually extracted storage and loss moduli),
is susceptible to provide more stringent tests for models.
Such experiments have been conducted recently:
because the flow was observed to remain homogeneous,
the measured behaviour can be robustly attributed to the local,
yet macroscopic, mechanical response of a 3D foam~\cite{Rouyer08}.
The strain-stress (Lissajous) curves display various shapes
and show that such data can become available
and provide non trivial results.
These results will be discussed later in the present work.

\subsection{Scope of the present work}

In materials whose plastic threshold corresponds to a small deformation,
shear banding requires a fluidizing mechanism 
such as those mentioned in paragraphe~\ref{SubSec:shb_complex_fluids} above.
But for materials whose deformation at plasticity onset is large, like foams,
the tensorial nature of the material state, due to stored deformation,
is sufficient to obtain shear bands: no extra mechanism is required.

In particular, the model that we suggest 
does not incorporate such an ingredient as dilatancy.
We know that a local plastic event results in an elastic redistribution 
of stress in the neighbourhood~\cite{falk_langer,Manning_Langer_Carlson_PRE_2007_76_056106}.
In simple shear geometry, it thus favours flow localization~\cite{kabla_debregeas,picard,picard_2005}.
In a statistical manner, it then raises locally the material fluidity~\cite{Bocquet_Colin_Ajdari_PRL_2009_103_036001}
and generates a {\em non-local} material rheology.
This non-local character had been observed in concentrated emulsions flowing 
in microfluidic channels~\cite{goyon_2008_454_84}.
Let us emphasize that these non-local effects are intrinsically present in our modelling
since the underlying elastic propagators~\cite{kabla_debregeas,picard,picard_2005,Bocquet_Colin_Ajdari_PRL_2009_103_036001} 
result directly from the elastic continuum medium equations that we use.

The paper is organized as follows.
We first discuss several types of continuum models for foams (Section~\ref{Sec:type_continuum_model}).
We then recall (Section~\ref{Sec:construction}) the construction of a rather generic continuum model 
for foam or emulsion rheology~\cite{benito_2008_225_elasto_visco_plastic_foam}.
It is generic as for the elasticity, the plastic flow rate
and the specifically three-dimensional form of the response.
We then simulate large amplitude oscillatory shear (Section~\ref{Sec:oscillatory_rheology})
and conduct a first round of comparison
with published data on such experiments at a fixed frequency for various amplitudes~\cite{Rouyer08}.
We are not able to reproduce the experimental data in a reasonable way with a single set of parameters,
but in the future, fitting similar data over a full range of both frequency and amplitude
will be a very effective and stringent method for testing more general models than the present one.
Finally, we thoroughly discuss shear banding in our model (Section~\ref{Sec:shb}).
Note that in this work, we restrain ourselves to a strictly mechanical and thermodynamical formulation. 
The important problems related to the coupling between the rheological behaviour 
and the structure of the material are not discussed. 
This coupling is experimentally well documented in various complex fluids systems 
in which shear bands are {\em de facto} associated with structural transitions~\cite{lerouge_berret_2010_AdvPolymSci}.

Here, we ask a more restricted question: 
could {\em stationary} shear bands in foams and emulsions be accounted for, 
starting from an inhomogeneous initial stress distribution 
in the material with otherwise strictly homogeneous mechanical properties?
Our main result: shear bands can emerge in a structurally homogeneous material under shear, 
only due to an inhomogeneous distribution of the initial internal stress in the material. 
We demonstrate this for a physically very natural form of the elastic and plastic laws.

	\begin{figure}
	\begin{center}
	\resizebox{0.8\columnwidth}{!}{%
	  \includegraphics{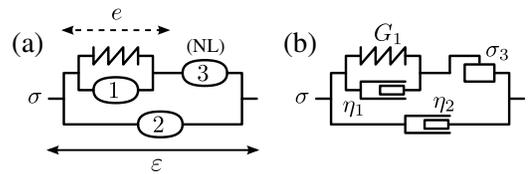}
	}
	\end{center}
	\caption{Schematic, scalar view of rheological models 
	(deformation $\varepsilon$ and stress $\sigma$) 
	with only one internal degree of freedom, 
	namely the deformation $e$ of the spring.
	{\em (a)} General model: it has one (non-linear) spring
	and three creeping elements, each of which may have a flow threshold.
	{\em (b)} Foams and emulsions display some (large) deformation
	before creep is triggered. Hence, creep elements~1 and~2
	cannot have a finite threshold.
	By contrast, element~3 does have a threshold.
	Here, for simplicity, we assume that the spring
	and both viscous elements are linear,
	and that element~3 cannot withstand any stress beyond $\sigma_3$.}
	\label{modele_general_une_variable_et_version_seuil}
	\end{figure}

	\begin{figure}
	\begin{center}
	\resizebox{0.7\columnwidth}{!}{%
	  \includegraphics{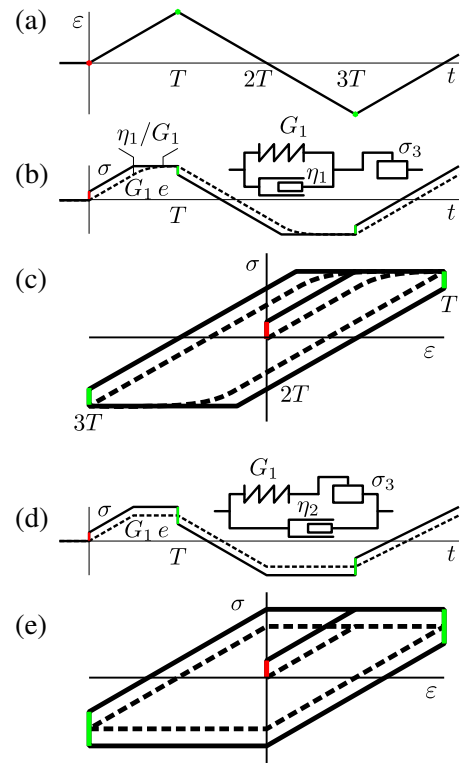}
	}
	\end{center}
	\caption{Response of the model represented in
	Fig.~\ref{modele_general_une_variable_et_version_seuil}b
	to a triangle wave deformation.
	{\em (a)} Deformation $\varepsilon$ as a function of time.
	{\em (b)} In the limit $\eta_2=0$, stress $\sigma$ (solid line)
	and elastic stress $G_1\,e$ transmitted by the spring (dashed line)
	as a function of time. When the threshold $\sigma_3$ is reached,
	the spring relaxes with the time scale $\eta_1/G_1$ of the Voigt element.
	{\em (c)} Corresponding Lissajous representation of $\sigma$ and $G_1\,e$.
	The periodic jump in stress (green segment)
	has the same amplitude as the initial jump (red segment).
	{\em (d)} In the limit $\eta_1=0$, stress $\sigma$ (solid line)
	and elastic stress $G_1\,e$ transmitted by the spring (dashed line)
	as a function of time.
	{\em (e)} Corresponding Lissajous representation of $\sigma$ and $G_1\,e$.
	The amplitude of the periodic jump in stress (green segment)
	is twice that of the initial jump (red segment).
	}
	\label{modeles_avec_un_seul_piston_et_lissajous}
	\end{figure}

\section{Choosing the type of continuum model}
\label{Sec:type_continuum_model}

As mentioned above, our main point is to explore models
{\it without} any extra dynamic variables
apart from the stored local deformation.
In the present Section, we will discuss these models in very general terms,
omitting any explicit tensorial features.

\subsection{Models with only one internal variable}
\label{models_with_only_one_internal_variable}

The most general arrangement of rheological elements
having only one internal deformation variable 
is represented in Fig.~\ref{modele_general_une_variable_et_version_seuil}a.
The deformation of the (not necessarily linear) spring
represents the deformation of the local structure.
Three creep elements (viscous and/or plastic) can be included, as shown.

At this point, let us recall that foams
are viscoelastic under weak stress conditions.
Hence, both the spring itself 
and the combination of spring and creep elements
must be able to deform under weak applied stress.
As a result, creep elements 1 and 2 must flow under weak stress:
we cannot choose them with a stress threshold 
below which they would not flow at all.
In other words, they are purely viscous (although not necessarily linear).
By contrast, creep element 3 must have a stress threshold
so that the entire system also displays a stress threshold.
These considerations are summarized 
schematically in Fig.~\ref{modele_general_une_variable_et_version_seuil}b.

Although elements 1 and~2 are viscous, 
they play different roles 
when some creep motion of element 3 is involved.
Let us first illustrate this point
by considering an experiment in which 
we impose a constant deformation rate from $t=0$
and reverse the deformation rate as of $t=T$.
For simplicity, we consider 
a linear spring with modulus $G_1$,
Newtonian viscosities $\eta_1$ and $\eta_2$.
Furthermore, we consider that element 3
is simply a solid friction element with threshold $\sigma_3$
with no dependence on velocity.
Fig.~\ref{modeles_avec_un_seul_piston_et_lissajous}
shows the contributions of viscous elements~1 and~2 separately
in the response of such a system to a triangle wave deformation.
In both cases, the stress jumps up immediately
to a finite value at $t=0$ due to the viscous elements $\eta_1$ and $\eta_2$.
The stress then increases at a constant rate as the spring elongates.
When the threshold of element~3 is reached, 
the stress saturates and remains constant.
At $t=T$, when the deformation rate is reversed,
the stress jumps down by a finite amount.
It then decreases at a constant rate as the spring
is relaxed and later stretched in the reverse direction.

There are two differences between the effects of viscous elements~1 and~2.
The first difference is that the observed threshold
depends on the deformation rate in the case of viscous element~2.
But that feature is not essential:
one can always decide that the deformation rate of creep element~3
affects its stress (in other words, by considering that
it contains not only a solid friction element,
but also an extra viscous element in parallel with it).

The second difference between both situations
of Fig.~\ref{modeles_avec_un_seul_piston_et_lissajous} is more essential.
The jumps in stress at $t=0$ and at $t=T$
have equal magnitudes in the case of viscous element~1.
By contrast, in the case of element~2,
the magnitude of the second jump is twice as large
as that of the initial jump 
(except if $T$ is too short for the system
to be able to relax the Voigt element,
with timescale $\eta_1/G_1$,
after it has reached the threshold stress).
More generally, in such an experiment,
one can express each viscosity 
(at the applied deformation rate)
in terms of the magnitudes of the stress jumps 
at $t=0$, when the deformation rate changes from zero to $+\gd$,
and at $t=T$, when it is reversed from $+\gd$ to $-\gd$:
\bee
\eta_1(\gd)&=&\frac{2|\Delta\sigma(0)|-|\Delta\sigma(T)|}{\gd}\\
\eta_2(\gd)&=&\frac{|\Delta\sigma(T)|-|\Delta\sigma(0)|}{\gd}
\eee
When the value of the elastic deformation $e$
can be measured independently,
for instance through optical measurements in 2D foams
and relevant statistical 
tools~\cite{graner_2008_369},
the respective contributions from elements~1 and~2
can be obtained by comparing $\sigma$ and $e$
(see full lines {\em versus} dashed lines
in Fig.~\ref{modeles_avec_un_seul_piston_et_lissajous}).

	\begin{figure}
	\begin{center}
	\resizebox{1.0\columnwidth}{!}{%
	  \includegraphics{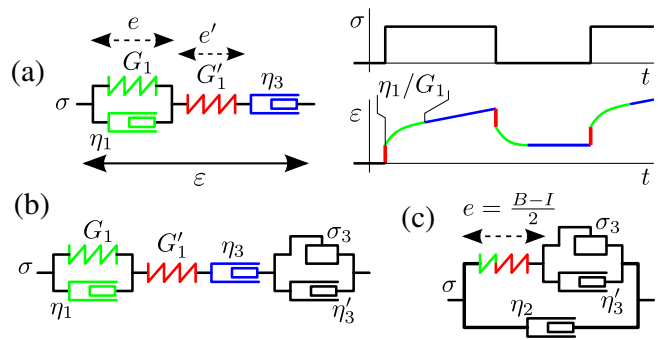}
	}
	\end{center}
	\caption{{\em (a)} Burger model: 
	schematic diagramme (left)
	and deformation response $\varepsilon(t)$
	to a step imposed stress $\sigma(t)$.
	Spring $G^\prime_1$ elongates immediately (red segment)
	while spring $G_1$ responds with some delay (green curve)
	due to viscous element $\eta_1$.
	Viscous element $\eta_3$ gives rise
	to a constant additional deformation rate (blue segment).
	{\em (b)} Combined Bingham-Burger model: 
	solid friction element $\sigma_3$
	(complemented by viscous element $\eta^\prime_3$)
	is now included so as to provide additional creep
	above the stress threshold.
	We believe that some tensorial version of this model
	could mimic the rheological behaviour of a foam quite adequately.
	{\em (c)} Model studied in the present work.
	Apart from the additional viscosity $\eta_2$
	introduced in Fig.~\ref{modele_general_une_variable_et_version_seuil},
	it represents the combined Bingham-Burger model
	when the (blue) viscous element has not been deformed yet,
	either at intermediate time scales 
	when the (green) Voigt element has relaxed (the green and red springs then respond in series)
	or at short time scales when the Voigt element is still blocked 
	(in which case only the red spring responds).}
	\label{modeles_Burger_et_deux_variables_et_le_notre}
	\end{figure}

\subsection{Weak stress: role of one extra internal variable}
\label{sec:weak_stress}

Below the flow threshold,
the model outlined above behaves 
like a Voigt element (a spring in parallel with a viscous element).
The behaviour of a foam under a weak stress
is in fact a little more complex: a linear Burger model 
(see Fig.~\ref{modeles_Burger_et_deux_variables_et_le_notre}a)
is known to correctly reproduce stepwise creep experiments 
on liquid foams~\cite{sylvie_cohen_addad_rh_yk_2004}.
Because the Burger model contains two springs,
it corresponds to a system with two internal variables
(spring elongations $e$ and $e^\prime$)
rather than one.
Fig.~\ref{modeles_Burger_et_deux_variables_et_le_notre}a 
shows the response of such a model to a stepwise creep experiment.
The stress jump generates an immediate elongation
of the (red) spring labeled $G^\prime_1$.
The (green) Voigt element then relaxes within a time scale
of order $\eta_1/G_1$.
On very long time scales, the (blue) viscous element
gives rise to a slow drift with velocity $\sigma/\eta_3$
(depicted by the finite slope of the blue line).

In order to build a model that behaves 
like the Burger model under weak stresses
but which presents a flow threshold
like discussed earlier,
one could combine the Burger model with the model presented 
in Fig.~\ref{modele_general_une_variable_et_version_seuil}
for only one internal variable.
We would thus obtain the model represented
in Fig.~\ref{modeles_Burger_et_deux_variables_et_le_notre}b
(which we already suggested as a generalization 
of our model~\cite{benito_2008_225_elasto_visco_plastic_foam}).

In the present work, we consider this combined model
of Fig.~\ref{modeles_Burger_et_deux_variables_et_le_notre}b,
but we focus on short and intermediate time scales
where the highly viscous (blue) element has not moved yet.
Because the blue element has not moved, it can be simply omitted.
As for the two springs $G_1$ and $G^\prime_1$
and the viscous element $\eta_1$,
their behaviour can be reduced to that of a single spring in two limits.
On time scales much shorter than $\eta_1/G_1$, 
the (green) Voigt element is blocked due to its viscous part.
As a result, the whole system behaves like
the red spring $G^\prime_1$.
At intermediate time scales (much longer than $\eta_1/G_1$
but still with a blocked blue viscous element),
the Voigt element has relaxed.
Hence, both springs are simply in series:
they combine into a composite spring.
In Fig.~\ref{modeles_Burger_et_deux_variables_et_le_notre}c,
we have represented such a model.
The green and red spring represents
either the red spring $G^\prime_1$ (on short time scales)
or the combination $G_1G^\prime_1/(G_1+G^\prime_1)$ 
(on intermediate time scales).
Meanwhile, the other creep elements provide 
both the stress threshold (solid friction $\sigma_3$)
and the dependence on deformation rate (viscous element $\eta^\prime_3$).
We also include a general viscosity $\eta_2$ as in the discussion 
of Paragraph~\ref{models_with_only_one_internal_variable}.
In this model, the spring and viscous elements
must be understood as non-linear, unless stated otherwise.
Apart from the viscous element $\eta_2$, the model
of Fig.~\ref{modeles_Burger_et_deux_variables_et_le_notre}c
is identical to the model that we constructed earlier
and for which we had analysed the local, mean field
behaviour~\cite{benito_2008_225_elasto_visco_plastic_foam}.

\section{Constructing the tensorial model}
\label{Sec:construction}

In the present Section, we will briefly recall
how we built the rheological model~\cite{benito_2008_225_elasto_visco_plastic_foam}
of Fig.~\ref{modeles_Burger_et_deux_variables_et_le_notre}c.
In particular, it is based on a general nonlinear description
of elasticity and plasticity. 
Indeed, materials such as foams 
can locally undergo large elastic deformations 
--- located far from the linear regime 
corresponding to small deformations --- before plastic flow 
occurs~\cite{weaire_hutzler_1999_book,french_book_belin_2010}.

\subsection{General local rheological laws}

The relevant framework to describe elastic stresses 
in a flowing material is the Eulerian one, 
whether this material possesses elastical properties or not. 
Indeed, during the flow of a foam or an emulsion, 
even though elastic stresses exist, 
any reminiscence of a {\em reference state} disappears continuously 
due to plasticity. 
The Lagrangian description, which is based on maintaining 
the correspondance with such an intial state of reference, 
is formally equivalent, 
but less adapted conceptually and numerically.

Thus we attach the variables describing the material 
to a spatial grid $(x,y,z)$, 
and they correspond to an instantaneous and local description 
in space.

In this framwork, only two variables are relevant 
in a strictly mechanical context: 
the local velocity gradient $\nabla\vec{v}(x,y,z)$ 
and the local deformation state stored in the 
material~\cite{benito_2008_225_elasto_visco_plastic_foam}
(green-red spring 
in Fig.~\ref{modeles_Burger_et_deux_variables_et_le_notre}c), 
as described in continuum mechanics 
by the Finger tensor $\Finger(x,y,z)$~\cite{bertram_2005}.
Note that when the material is at rest, 
$\Finger=\unittensor$ while the stored deformation,
depicted schematically in Fig.~\ref{modeles_Burger_et_deux_variables_et_le_notre}c, 
vanishes: $e=0$.

In this section, we describe our local rheological model. 
Note that in this local context, the global tensor $\gradv$ itself
has to be considered as an independent local three-dimensional 
tensorial variable, just as $\Finger$, 
not as the spatial gradient of a velocity field. 
Only when we will turn to the description of a spatial system, 
see section~\ref{Sec:regime_ecoul_inhomog_stat}, 
will the vector field $\vec{v}(x,y,z)$ be introduced. 
Meanwhile, tensors $\gradv$ and $\Finger$ 
will thus be the two variables of our local tensorial model.

The elastic part of the stress, which goes throuth the spring
in Fig.~\ref{modeles_Burger_et_deux_variables_et_le_notre}c
depends on the deformation according to the following relation, 
the most general one compatible with the symmetry constraints 
in three dimensions~\cite{benito_2008_225_elasto_visco_plastic_foam}: 
\be
\label{elasticLawIBB2_0}
\si_{\rm el}=a_0\,\unittensor+a_1\,\Finger+a_2\,\Finger^2,
\ee
where $a_0$, $a_1$ and $a_2$ are scalar functions 
of the invariants of the Finger tensor $\Finger$.

Turning to plasticity
($\sigma_3$ and $\eta^\prime_3$
in Fig.~\ref{modeles_Burger_et_deux_variables_et_le_notre}c), 
we only assume 
that every event of plastic relaxation is {\em aligned} 
with the stored deformation. 
The plastic creep $\epspp$ should thus be similarly aligned. 
The most general form compatible 
with the symmetry constraints is then:
\be
\label{epspp_decomp_Finger_sans_dev}
\epspp=\bar{b}_0\,\unittensor
+\bar{b}_1\,\Finger
+\bar{b}_2\,\Finger^2,
\ee
where $\bar{b}_0$, $\bar{b}_1$ and $\bar{b}_2$ 
are again scalar functions of the invariants 
of the Finger tensor $\Finger$.

To complete the model, 
we gather together in a global viscosity term 
(which was noted $\eta_2$
in Fig.~\ref{modeles_Burger_et_deux_variables_et_le_notre}c)
all the dissipative phenomena which are present 
even in the absence of any plastic event in the foam. 
They occur for example at small scales: 
flows in films squeezed between bubbles or in Plateau borders.
We simplify its description in selecting 
a Newtonian average viscosity $\eta_s$
for these local dissipative phenomena. 
The list of contributions to the stresses 
in the material is thus closed. We have:
\be
\label{elasticLawPlusViscosity}
\si=a_0\,\unittensor+a_1\,\Finger+a_2\,\Finger^2
+\frac{\eta_s}{2}(\gradv+\transp{\gradv}).
\ee

To take into account the incompressible character 
of foams and emulsions, 
we add an extra kinematic constraint of strict volume conservation
$\det(\Finger)=1$.
Refering to \cite{benito_2008_225_elasto_visco_plastic_foam} 
for further details, we take it into account
by using only the deviatoric part of the stress:
\be
\s=\dev(\si)=\si-\frac{\unittensor}{d}\trace(\si).
\ee
The same constraint on plasticity gives 
the general form~\cite{benito_2008_225_elasto_visco_plastic_foam}:
\be
\label{epspp_decomp_Finger}
\epspp=\Finger\cdot\dev(f(\Finger))=b_1\,\Finger\cdot\dev(\Finger)
+b_2\,\Finger\cdot\dev(\Finger^2),
\ee
where the scalar prefactors $b_1$ and $b_2$ are isotropic, 
and thus depend on the invariants of tensor $\Finger$.

In what follows, we will use a completely equivalent form 
of tensor $\epspp$ which manifests more clearly 
that the dissipation is positive
(see the discussion in~\cite{benito_2008_225_elasto_visco_plastic_foam}):
\be
\label{epspp_decomp_positivity}
\epspp=\frac{\amplepspp(\Finger)}{\tau}\,
\Finger\cdot{\cal G}(\Finger)
\ee
where $\amplepspp(\Finger)$ 
is a scalar isotropic function of $\Finger$, 
$\tau$ the characteristic time of the dissipative processes; 
moreover:
\be
\label{eq:epspphat}
{\cal G}(\Finger)
=\frac{\dev\left[{\cal P}(\Finger)\cdot\dev(\si_{\rm el})\right]}
{\trace\left[{\cal P}(\Finger)\cdot\dev(\si_{\rm el})\cdot\dev(\si_{\rm el})\right]},
\ee
with ${\cal P}$ is a function of the form ${\cal P}(\Finger)=b(\Finger)\, \Finger^{-2}+(1-b(\Finger))\,\Finger^2$~\cite{benito_2008_225_elasto_visco_plastic_foam},
where $b$ is an isotropic function.
In this expression, the total dissipation 
per unit volume is $\amplepspp(\Finger)$
and can be chosen as positive.

Eventually one gets the generic local rheological model:
\bee
&&\frac{{\rm d}\Finger}{{\rm d}t}
-\gradv \cdot \Finger
-\Finger \cdot \transp{\gradv}
=-2\,\epspp,\\
&&\epspp=\frac{\amplepspp(\Finger)}{\tau}\,
\Finger\cdot{\cal G}(\Finger),\\
&&\si=a_0\,\unittensor+a_1\,\Finger+a_2\,\Finger^2
+\frac{\eta_s}{2}(\gradv+\transp{\gradv}),
\eee
where ${\rm d}\Finger/{\rm d}t
=\partial\Finger/\partial t
+(\vec{v}\cdot\nabla)\Finger$
is the particulate derivative of the Finger tensor.

\subsection{Complete spatial model}

As for any local rheological model, 
the previous equations must be complemented 
by field equations which express force balance 
and mass conservation:

\bee
\label{force_balance_equation}
\nabla\cdot\s+\rho\,\vec{f}
=\rho\frac{{\rm d}\vec{v}}{{\rm d}t}
-\vec{\nabla}\,p,\\
\label{density_evolution}
\frac{\partial\rho}{\partial t}
+\nabla\cdot(\rho\;\vec{v})
=\frac{{\rm d}\rho}{\dt}
+\rho\,\trace\frac{1}{2}(\gradv+\transp{\gradv})
=0,
\eee
where $\vec{f}$ represents the external forces (per unit mass), 
and $\rho$ is density. 
The incompressibility constraint gives here
\be
\nabla \cdot \vv=\trace\frac{1}{2}(\gradv+\transp{\gradv})=0.
\ee
As a result, the density $\rho$ is simply transported by the flow:
${\rm d}\rho/\dt=0$.
In the remaining of this work,
we furthermore assume that the density is homogeneous,
hence it also remains constant: $\partial\rho/\partial t=0$.

Last assumption: we restrict ourselves to the Stokes regime, 
where inertial terms are all negligible 
in the mass conservation equation. Thus one obtains:
\be
\nabla\cdot\s=-\vec{\nabla}\,p.
\ee
The complete system of equations that we need
to integrate numerically is thus:
\bee
\label{eq:stokescomplet}
&&\frac{{\rm d}\Finger}{{\rm d}t}
-\gradv \cdot \Finger
-\Finger \cdot \transp{\gradv}
=-2\,\epspp,\\
&&\epspp=\frac{\amplepspp(\Finger)}{\tau}\,
\Finger\cdot{\cal G}(\Finger),\\
\label{Eq:trace_grad_v_zero}
&&\trace\frac{1}{2}(\gradv+\transp{\gradv})=0,\\
\label{Eq:sigma_dev}
&&\s=\dev(\si)\nonumber\\
&&\,\,\,\,\,=\dev\left\{a_1\,\Finger+a_2\,\Finger^2\right\}
+\frac{\eta_s}{2}(\gradv+\transp{\gradv}),\\
\label{Eq:div_sigma}
&&\nabla\cdot\s=-\vec{\nabla}\,p,
\eee

The initial conditions that must be specified
to solve the above system
may merely consist in the values of tensor $\Finger$
over the entire sample.
Indeed, the value of the velocity and pressure fields
can be derived therefrom using
equations~(\ref{Eq:div_sigma}) and~(\ref{Eq:sigma_dev})
which, when combined, are similar to Stokes' equation,
using the constraint of equation~(\ref{Eq:trace_grad_v_zero}).

\subsection{Selection of a particular form of elasticity and plasticity}

\subsubsection{Elasticity: Mooney-Rivlin model}

We have selected  a usual form of incompressible elasticity 
which has been demonstrated to describe 
to a good approximation the nonlinear elastic behaviour 
of foams~\cite{hoehler_cohen_addad_review_2005,hoehler_cohenaddad_labiausse_2004}: 
Mooney-Rivlin elasticity.
The corresponding elastic energy per unit volume
can be written~\cite{bertram_2005}:
\be
\rho\,\W(\Finger)=\frac{\mra}{2}({\rm I}_\Finger-3)
+\frac{\mrb}{2}({\rm II}_\Finger-3)
\ee
where
\bee
{\rm I}_\Finger&=&\trace(\Finger)\\
{\rm II}_\Finger&=&\frac12[\trace^2(\Finger)-\trace(\Finger^2)]
=\trace(\Finger^{-1})
\eee
Going back the coeficients of Eq.~(\ref{elasticLawIBB2_0}),
this corresponds to the following expressions:
\bee
\label{a1_a2_mooney_rivlin}
a_1&=&\mra+\mrb\,{\rm I}_\Finger\\
a_2&=&-\mrb
\eee
Following previous work refs.~\cite{hoehler_cohen_addad_review_2005,hoehler_cohenaddad_labiausse_2004}, we express the values of $\mra$ and $\mrb$ using an elastic modulus $\G$
and an interpolation parameter $a$ as follows:
\bee
\label{eq:moonleyrivlink1}\mra&=&a\G\\
\label{eq:moonleyrivlink2}\mrb&=&(1-a)\G.
\eee
In the foam modelling litterature, a value $a=1/7$ 
is sometimes recommended~\cite{hoehler_cohen_addad_review_2005,hoehler_cohenaddad_labiausse_2004}. 
Keeping in mind our perspective of discussing 
the conditions for the appearance of shear bands 
depending on parameter values, 
in sections~\ref{impact_elasticity_plasticity} and beyond, 
we prefer to keep the parameter $a$ free, 
although we remain in the framework of the Moonley-Rivlin elasticity.

\subsubsection{Plasticity: yield stress fluid}

The particular form of plasticity explored in this work is based on a nonlinear threshold-like behaviour. 
Locally, the plastic reorganisation events 
only occur in the material 
when the stored elastic deformation 
reaches a critical value.
We express this transition with a function 
$\WW_y(\Finger)$ which vanishes linearly at the threshold:
\be
\label{threshold_function}
\WW_y(\Finger)=0,
\ee
with, in our case, $\WW_y(\Finger)=\rho\W(\Finger)-K$,
where $\rho\W$ is the stored elastic energy per unit volume,
and $K$ a constant.
In simple shear from a relaxed state,
$\si_y$ is the threshold stress:
function $\WW_y$ vanishes.

{From} the point of view of the plastic 
deformation rate tensor $\epspp$, 
we have the following 
expression~\ref{epspp_decomp_positivity}, 
taking for $\amplepspp(\Finger)$:
\bee
\label{forme_Dp}
\amplepspp(\Finger)=(\rho\W(\Finger)-K)\, \Theta(\rho\W(\Finger)-K),
\eee
where $\Theta(x)=1$ when $x\geq 0$ and $\Theta(x)=0$ elsewhere. 

We also set the following form for the polynom: 
\be
\label{eq:polynome_P_b}
{\cal P} (\Finger) = b \Finger ^{-2}+(1-b) \Finger ^2.
\ee
with $b$ between 0 and 1.
Our final set of equations is thus:
\bee
\label{kinematic_final}
&&\frac{{\rm d}\Finger}{{\rm d}t}
-\gradv \cdot \Finger
-\Finger \cdot \transp{\gradv}
=-2\,\epspp,\\
\label{plasticity_final}
&&\epspp=\frac{\rho\W(\Finger)-K}{\tau}\, \Theta(\rho\W(\Finger)-K)\,\Finger\cdot{\cal G}(\Finger),\\
\label{si_final}
&&\s=\dev(\si)=\dev\left\{\left(a\,\G\,+\,(1-a)\,\G\,\trace(\Finger) \right)\,\Finger\right.\nonumber\\
	&&-\,(1-a)\G\,\Finger^2
	\left.+\eta_s(\gradv+\transp{\gradv})/2\right\},\\
\label{eq:divsigmagradp}
&&\nabla\cdot\s=-\vec{\nabla}p,\\
&&\trace\frac{1}{2}(\gradv+\transp{\gradv})=0.
\eee

\subsubsection{Physical parameters and rheological model}
\label{rheological_model}
In order to be able to highlight physically relevant quantities,
we use a non-dimensional form of the above system.
The elastic modulus $\G$ is taken as the unit of stress,
and the weak stress relaxation timescale $\eta_s/\G$
as the unit of time, while $\Finger$ is already dimensionless:
\bee
\hat{\si}&=&\s/\G,\\
{\cal T} &=& \eta_s\, t/\G,\\
\hat{\Finger}&=&\Finger.
\eee
As a result, the various quantities are non-dimensionalized as follows:
\bee
\hat{E}(\hat{\Finger})&=&\rho E(\Finger)/\G,\\
{\cal K}&=&K/\G,\\
\hat{p}&=&p/\G,\\
\hat{\cal A}(\Finger)&=&{\cal A}(\Finger)/\G,\\
\hat{\cal P}(\Finger)&=&{\cal P}(\Finger),\\
\hat{\cal G}(\Finger)&=&\G\,{\cal G}(\Finger),\\
\hat{\gradv}&=&(\eta_s/\G)\,\gradv,\\
\hat{\epspp}&=&(\eta_s/\G)\,\epspp.
\eee

The system of equations now reads:

\bee
\label{kinematic_final_adim}
&&\frac{{\rm d}\hat{\Finger}}{{\rm d}{\cal T}}
=\hat{\gradv} \cdot \hat{\Finger}
+\hat{\Finger} \cdot \transp{\hat{\gradv}}
-2 \hat{\epspp},\\
&&\trace(\hat{\gradv}+\transp{\hat{\gradv}})=0,\\
\label{eq:divsigmagradp_adim}
&&\nabla\cdot\hat{\si}=-\vec{\nabla}\hat{p},\\
\label{si_final_adim}
&&\hat{\si}=\hat{\siel}+\frac{\hat{\gradv}+\transp{\hat{\gradv}}}{2},\\
\label{eq:elast_dev_sigmael_adim}
&&\hat{\siel}=\left(a\,+\,(1-a)\,\trace(\Finger) \right)\,\dev\hat{\Finger}\nonumber\\
&&\hspace{1cm}-\,(1-a)\,\dev\hat{\Finger}^2,\\
\label{plasticity_final_adim}
&&\hat{\epspp}=\Psi\,\hat{\amplepspp}(\hat{\Finger})\,\hat{\Finger}\cdot\hat{{\cal G}}(\hat{\Finger}),\\
&&\hat{\amplepspp}(\hat{\Finger})=(\hat{\W}(\hat{\Finger})-\hat{K})
\,\Theta(\hat{\W}(\hat{\Finger})-\hat{K}),\\
&&\hat{E}(\hat{\Finger})
=\frac{a}{2}({\rm I}_\Finger-3)+\frac{1-a}{2}({\rm II}_\Finger-3),\\
&&\hat{{\cal G}}(\hat{\Finger})
=\frac{\dev\left[{\cal P}(\Finger)\cdot\hat{\siel}\right]}
{\trace\left[{\cal P}(\Finger)\cdot\hat{\siel}\cdot\hat{\siel})\right]},\\
\label{eq:plast_function_P}
&&\hat{{\cal P}} (\hat{\Finger}) = b \hat{\Finger}^{-2}+(1-b) \hat{\Finger}^2\\
&&\Psi=\frac{\eta_s}{\G\,\tau}.
\eee

The new non-dimensional parameter $\Psi$
defined in the last equation above reflects the ratio 
of the plastic flow rate (proportional to $1/\tau$)
to the viscoelastic flow rate (proportional to $G/\eta_s$)
when the other factors have the same order of magnitude.
With the present choice for the magnitude of $\epspp$
(with $\hat{\amplepspp}$ proportional to the distance from the threshold),
that occurs when the stored deformation
is typically twice the threshold deformation.

\subsection{Simple shear flow}

In the remaining of this work,
we address specifically the question of shear banding.
For this purpose, we consider only simple shear flows.
The velocity is oriented along axis $x$
and varies along along axis $y$.
The only non-zero component of the velocity gradient $\nabla\vec{v}$
is then $\partial v_x/\partial y$.
The entire system and flow are invariant along $x$ and $z$.
Besides, the force balance given by Eq.~(\ref{eq:divsigmagradp})
then implies that $\sigma_{xy}$ and $\sigma_{yy}$
are homogeneous at all times.
In the axes $x$, $y$ and $z$, the non-dimensionalized velocity gradient
can thus be written as:
\be
\label{Eq:gradv_adim_y}
\hat{\gradv}=\begin{pmatrix} 0 & \dot{\Gamma}(y) & 0 \\ 0 & 0 & 0 \\ 0 & 0 & 0 \end{pmatrix}
\ee
where $\dot{\Gamma}(y)=(\eta_s/\G)\,\dot{\gamma}(y)$.
Let $\dot{\Gamma}=(\eta_s/\G)\,\dot{\gamma}$ 
be the macroscopic value of the shear rate
at the scale of the entire sample.
We now have five non-dimensional parameters:
the plastic-to-viscoelastic flow rate ratio $\Psi$,
the threshold ${\cal K}$, 
the Mooney-Rivlin parameter $a$, 
the parameter $b$ (which defines the tensorial form 
of the plasticity ${\cal G}(\Finger)$),
and the macroscopic shear rate $\dot{\Gamma}$.

For the sake of consistency,
let us note that our earlier work~\cite{these_sylvain_benito_2009}
discussed parameters
\bee
\alpha=\frac{2}{\dot{\Gamma}}\\
\weiss=\frac{\dot{\Gamma}}{\Psi}
\eee
instead of $\Psi$ and $\dot{\Gamma}$.

\subsection{Relation between model parameters and experimentally measurable quantities}

	\begin{figure}[htbp]
	\begin{center}
	\resizebox{0.8\columnwidth}{!}{%
	\includegraphics{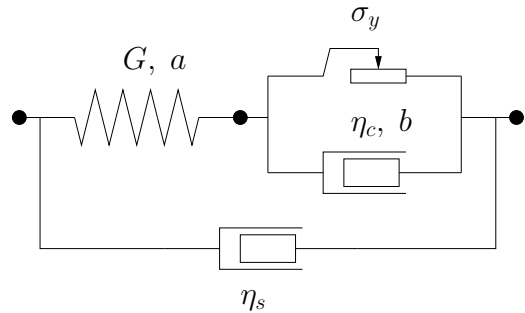}
	}
	\end{center}
	\caption{Simplified (scalar) picture
	of the main rheological parameters.
	$G$ represents the elastic modulus
	and $a$ the relative weight
	of the tensorial components of the elastic deformation,
	see Eqs.~(\ref{eq:moonleyrivlink1}) 
	and~(\ref{eq:moonleyrivlink2}). 
	The quantities $\sigma_y$ and $1/\tau$ 
	constitute a scalar representation
	of the creep defined by $\epspp$,
	and parameter $b$ is the equivalent of $a$ for creep,
	see equation~(\ref{eq:polynome_P_b}). 
	Finally, $\eta_s$ is a viscosity
	that is independent of creep.}
	\label{fig:ressort_friction_piston_avec_piston}
	\end{figure}
In fig.~\ref{fig:ressort_friction_piston_avec_piston}, we summarize
the physical parameters included in our model. 

In experiments, easily accessible dimensional parameters are 
the viscosity $\eta_s$ and the shear modulus $G$ 
through linear rheology, as well as the threshold stress $\sigma_y$. 
More elaborate setups can yield 
the value of $a$. There are indications that a value $a=1/7$
is relevant for liquid foams~\cite{hoehler_cohen_addad_review_2005,hoehler_cohenaddad_labiausse_2004}.

Among our non-dimensional parameters, two can thus be determined easily: $a$ and ${\cal K}$.
The latter is related to the energy at the threshold
${\cal K} \approx \frac{1}{2}(\sigma_y/G)^2$. 
As just mentionned $\dot{\Gamma}(y)=(\eta_s/\G)\,\dot{\gamma}(y)$ is the normalised shear rate.
Concerning $\Psi=\eta_s/(\G\,\tau)$ and $b$, no experiment to our knowledge
is able to validate or invalidate 
the value of the plastic reorganisation time $\tau$ at deformations beyond the threshold,
or the tensorial form of the plastic flow (here expressed in terms of parameter $b$).
For the time being, we thus consider parameters $\Psi$ and $b$ as free parameters in any 
comparison of our model with actual data.

\section{Homogeneous flow behaviour in large amplitude oscillatory experiments}
\label{impact_elasticity_plasticity}
\label{Sec:oscillatory_rheology}

	\begin{figure}[htbp]
	\begin{center}
	\begin{psfrags}
	\psfrag{Gamma}{$\Gamma_0$}
	\psfrag{omega}{$\omega$}
	\includegraphics[width=1\columnwidth]{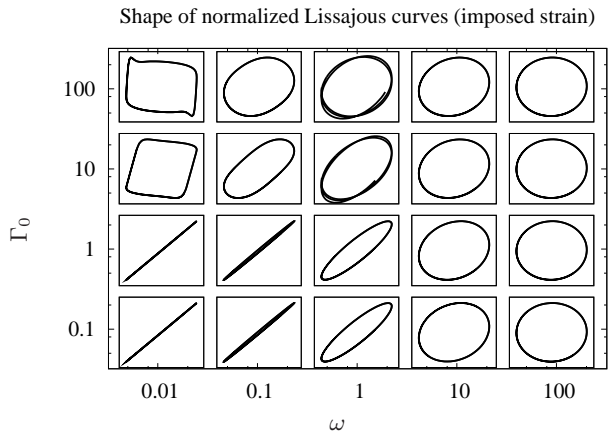}
	\end{psfrags}
	\end{center}
	\begin{center}
	\begin{psfrags}
	\psfrag{Gamma}{$\Gamma_0$}
	\psfrag{omega}{$\omega$}
	\includegraphics[width=1\columnwidth]{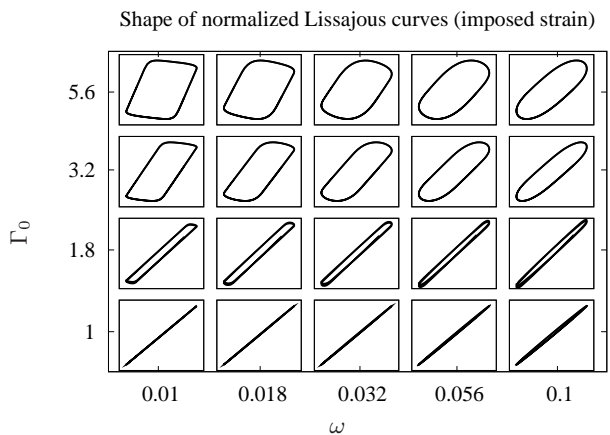}
	\end{psfrags}
	\end{center}
	\caption{Large amplitude oscillatory simulations: 
	shape of $\Gamma(t)$ {\em versus} $\sigma_{xy}(t)$ (Lissajous) curves,
	obtained for $a=b=1/7$, $\Psi=0.1$ and ${\cal K}=1$.
        Top: range from $\omega=0.01$ to $\omega=100$
        and $\Gamma_0=0.1$ to $\Gamma_0=100$.
        Bottom: zoom on a more restricted range of parameters.}
	\label{Fig:table_Lissajous_Gamma_nu}
	\end{figure}

	\begin{figure}[htbp]
	\begin{center}
	\begin{psfrags}
	\psfrag{Gamma / Gamma-max}{$\Gamma(t)/\Gamma_0$}
	\psfrag{Sigma / Sigma-max}{$\sigma(t)/\sigma_{\rm  max}$}
	\psfrag{omega = 0.01}{$\omega=0.01$, $\Gamma_0=1$ to $100$}
	\psfrag{omega}{$\omega$}
	\psfrag{Gamma=1}{$\Gamma_0=1$}
	\psfrag{Gamma=3.2}{$3.2$}
	\psfrag{Gamma=10}{$10$}
	\psfrag{Gamma=100}{$100$}
	\includegraphics[width=1\columnwidth]{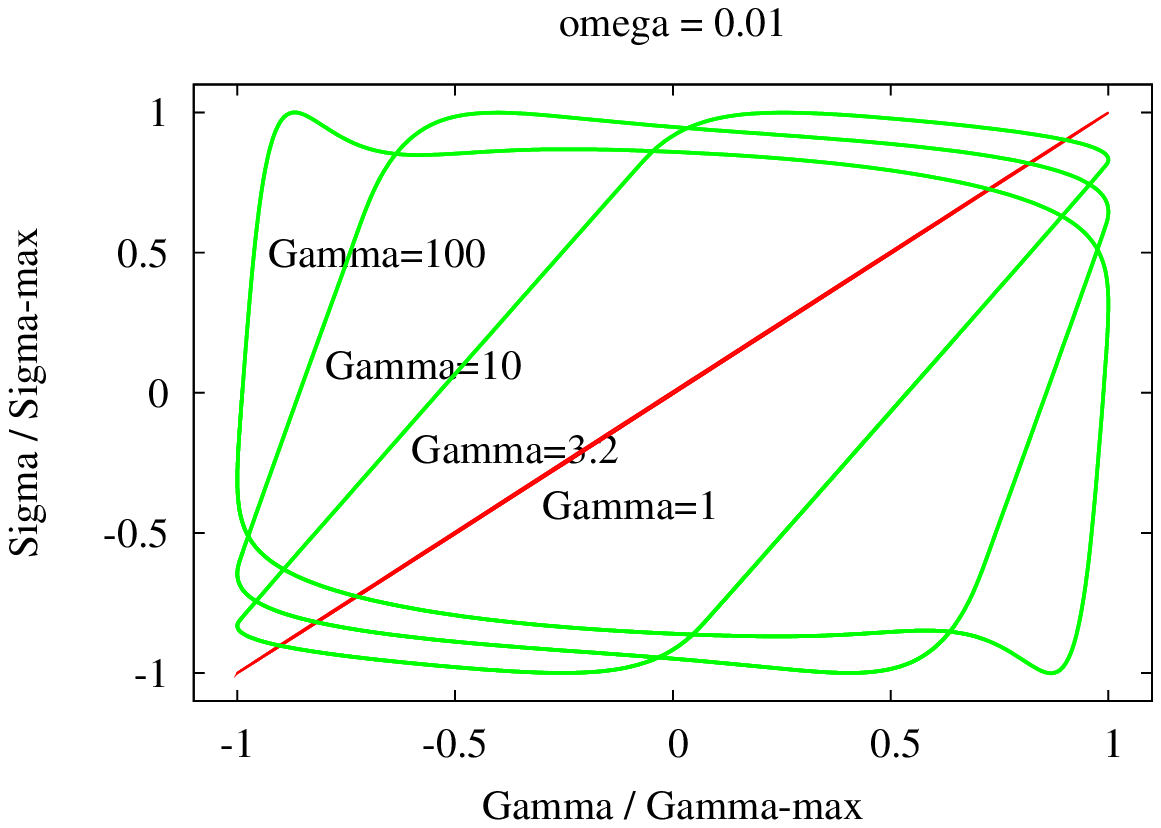}
	\end{psfrags}
	\end{center}
	\begin{center}
	\begin{psfrags}
	\psfrag{Gamma}{$\Gamma(t)$}
	\psfrag{Gamma=1.4}{$\Gamma_0=1.4$}
	\psfrag{Gamma=1.8}{$\Gamma_0=1.8$}
	\psfrag{Sigma}{$\sigma(t)$}
	\psfrag{omega = 0.01}{$\omega=0.01$, $\Gamma_0=1.4$ and $1.8$}
	\includegraphics[width=1\columnwidth]{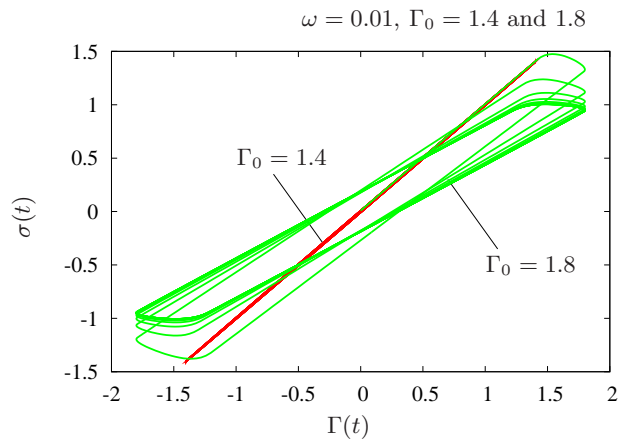}
	\end{psfrags}
	\end{center}
	\caption{Large amplitude oscillatory simulations: 
	shape of $\sigma_{xy}(t)$ {\em versus} $\Gamma(t)$ (Lissajous) curves,
	obtained for $a=b=1/7$, $\omega=0.01$, $\Psi=0.1$ and ${\cal K}=1$.
Top: transition from elastic to plastic behaviour as the amplitude is increased.
At very large amplitudes, an overshoot is apparent like in continuous shear situations.
Bottom: in a slightly plastic situation (amplitude 1.8),
it takes several cycles before the system behaves in a periodic manner.}
	\label{Fig:table_Lissajous_Gamma_nu2}
	\end{figure}

	\begin{figure}[htbp]
	\begin{center}
	\begin{psfrags}
	\psfrag{a=0.14 Kappa=0.02 b=0.14 Psi=5 omega=0.389}{Normalized stress over one period}
	\psfrag{stress + arbitrary constant}{$\sigma+{\rm const}$}
\includegraphics[width=1\columnwidth]{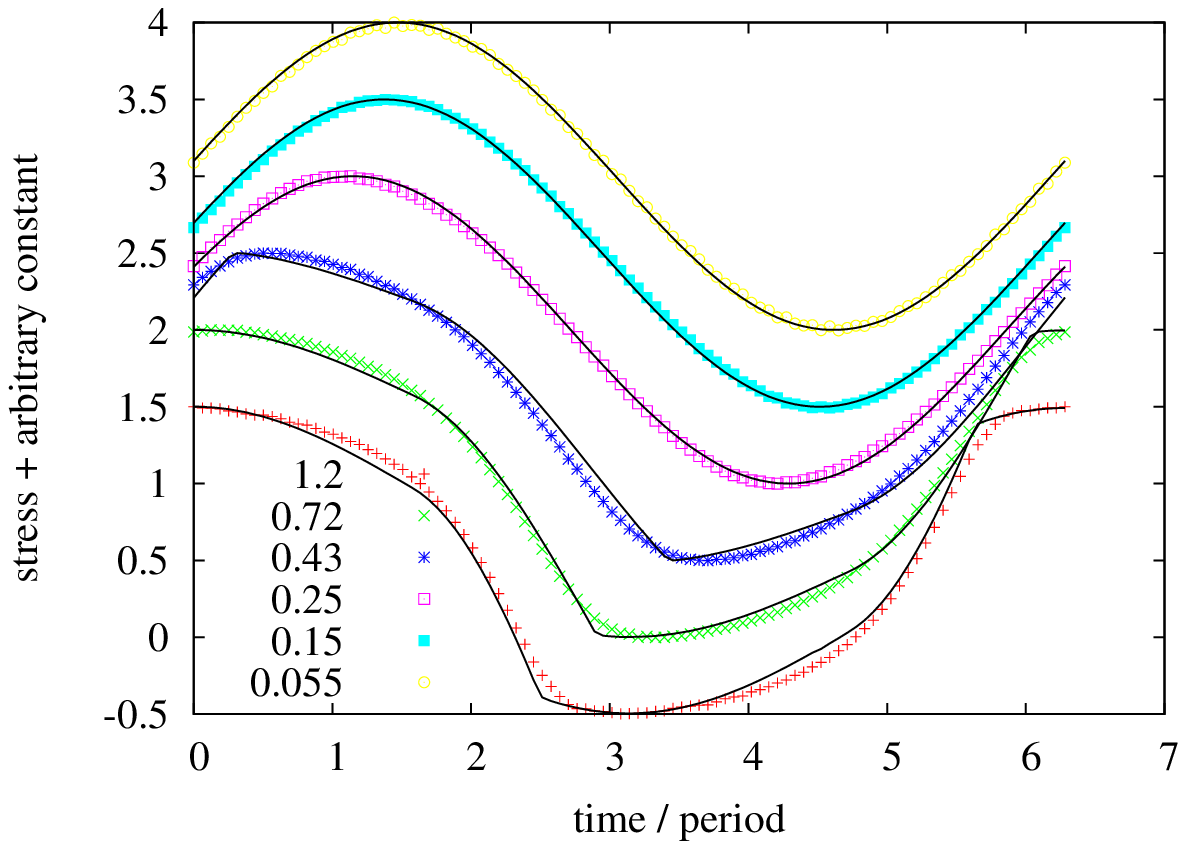}
	\end{psfrags}
	\end{center}
	\caption{Comparison of model (curves) with experiments (points). 
	Normalized $\sigma(t)$ curves 
for six values of the amplitude $\Gamma_0$ ranging from 0.055 to 1.2,
shifted vertically for clarity.
Parameter values are $a=0.14$, ${\cal K}=0.04$, $b=0.14$, $\Psi=27$.
For each value of the amplitude $\Gamma$ we had to select a different value for the frequency $\omega$ and for the modulus $G$.
The ($\Gamma$, $\omega$, $G$) values are:
(0.055, 0.2, 282), 
(0.15, 0.4, 246), 
(0.25, 0.9, 186),
(0.43, 0.5, 154),
(0.72, 0.5, 146),
(1.2, 0.4, 149).
}
	\label{Fig:fit_rouyer_normalized}
	\end{figure}

\subsection{Method}

In the present section, we test the predictions of our model
by comparing them to the most stringent available measurements
in homogeneous flows,
namely the large amplitude oscillatory experiments
conducted recently by Rouyer {\em et al.}~\cite{Rouyer08}.
We have simulated oscillatory shear flow
with the local model (no dependence on coordinate $y$,
{\em i.e.}, homogeneous flow).
In other words, in Eq.~(\ref{Eq:gradv_adim_y}), we choose
\be
\dot\Gamma(y,t)=\dot\Gamma(t)
=-\omega\Gamma_0\,\cos(\omega\,t),
\ee
which corresponds to the oscillating shear deformation:
\be
\Gamma(y,t)=\Gamma(t)
=\Gamma_0\,\sin(\omega\,t),
\ee

\subsection{Typical behaviours}

Figs.~\ref{Fig:table_Lissajous_Gamma_nu}
and~\ref{Fig:table_Lissajous_Gamma_nu2}
show, in the form of Lissajous curves,
the full response of the present model
in the (amplitude, frequency)-plane.
The normalized stress and strain responses
make it easy to discriminate between plastic, elastic, or viscous behaviours
of the model.
Let us rationalize them in terms of the scalar diagram
of Fig.~\ref{modeles_Burger_et_deux_variables_et_le_notre}c
discussed above in Section~\ref{sec:weak_stress}.

A pure elastic behaviour corresponds
to an ellipse squeezed into a straight line spanning the diagonal of the diagram.
This is obtained at low frequencies and amplitudes.
Indeed, deformation rates are then small at all times,
hence the viscous elements 
in diagram~\ref{modeles_Burger_et_deux_variables_et_le_notre}c
play no role.
Meanwhile, because deformations remain small,
the threshold of the solid friction element is never reached.
As a result, the spring alone provides the mechanical response of the system.

For the same reason, an elasto-plastic behaviour 
is expected at low frequency yet large amplitude,
since the stress threshold is then reached.
A purely elasto-plastic behaviour, as predicted by a scalar model,
would correspond to a sharp-cornered parallelogram
with two horizontal sides corresponding to the yield stress.
The results of our simulation at low frequency and large amplitude
differ from this simple picture
in the same way as experimental data 
by Rouyer {\em et al.}~\cite{Rouyer08},
namely with two main features:
{\em (i)} the ``plastic part'' of the Lissajous curve
exhibits a slightly negative slope,
and {\em (ii)} the transition to plasticity
is progressive rather than sharp (blunt corners).
Feature {\em (i)} corresponds to the weakening of the viscous component
when the deformation rate decreases
along the sinusoidal applied deformation.
As for the latter feature, it can result either 
{from} viscosity being combined with plasticity
(as in the present model~\cite{benito_2008_225_elasto_visco_plastic_foam})
or from a progressive onset of plasticity~\cite{marmottant_graner}.

As compared to the results by Rouyer {\em et al.}~\cite{Rouyer08},
our model additionally exhibits an overshoot
at very low frequency and large amplitude.
Because such a regime is very similar to slow, continuous shear,
this response can be understood~\cite{raufaste_2010_0732} 
as a tensorial effect
combining the saturation arising from plasticity 
and the rotation contained in shear
(this point is further discussed at the end of Section~\ref{Sec:large_eldef_dyn_var}).

This transition between a purely elastic response at low amplitude 
and an elasto-plastic response at higher amplitude
is best illustrated by the top part of Fig.~\ref{Fig:table_Lissajous_Gamma_nu2}.
The curves are normalized for clarity,
but the actual maximum slope in each curve
is essentially identical and is given by the shear modulus $\G$.
By contrast, the value of the stress in the most horizontal regions of the curve
reflect both the solid friction element and both viscous elements
in diagram~\ref{modeles_Burger_et_deux_variables_et_le_notre}c.
The bottom part of Fig.~\ref{Fig:table_Lissajous_Gamma_nu2}
presents results slightly below and slightly above the plasticity threshold.
It shows that weak plasticity causes the deformation cycle 
to slowly drift towards a limit cycle that differs from the elastic cycle.
This very drift, when continued along a longer cycle,
is in fact at the origin of the overshoot mentioned above.

If we now turn to higher frequencies,
the deformation rate becomes large.
As a result, viscous elements now play a role
and may even become dominant.
Correspondingly, the Lissajous curve
tend to become an ellipse whose axes lie along those of the figure.
That is particularly clear on diagram~\ref{modeles_Burger_et_deux_variables_et_le_notre}c
at high frequency with a large amplitude,
but the trend is very obvious at high frequency and low amplitude,
and is also discernable at low frequency and large amplitude.

\subsection{Comparison with experiments}
\label{sec:comparison_with_experiments}

The data obtained by Rouyer {\em et al.}~\cite{Rouyer08}
correspond to a fixed frequency and different applied strain amplitudes (from $0.055$ to $1.2$).
We have integrated%
~\footnote{Note that the {\tt OCTAVE} software code for our 
model simulation is freely available on our website}
the equations of the present model
with the same amplitudes and plotted them together with data.
The results are presented in figure~\ref{Fig:fit_rouyer_normalized}
(for clarity, stress curves are presented normalized).

Note that in order to obtain a reasonable agreement of our model with the data,
we had to artifically choose different values of $\omega$ and $\G$
for each strain amplitude,
while $\kappa$ and $\Psi$ could be kept constant.
This unsatisfying {\em ad hoc} parameter adjustment
shows the limits of this model
in describing the behaviour of the foams
studied by Rouyer {\em et al.}

\section{Shear banding study}
\label{Sec:shb}

	\begin{figure}[ht!]
	\begin{center}
	\resizebox{0.8\columnwidth}{!}{
	\includegraphics{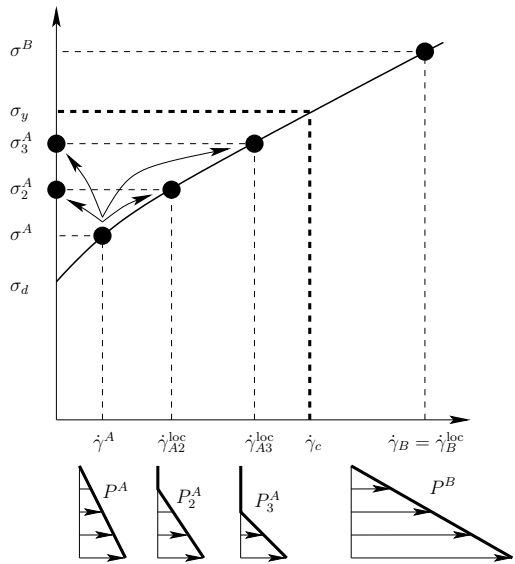}
	}
	\end{center}
	\caption{Typical form of a stationnary flow curve 
	giving the dependence of the shear stress on the local shear rate.  
	$\sigma_y$ is the yield stress as measured 
	under imposed stress, and  $\gd_c$ the corresponding shear rate.
	A macroscopic shear rate $\gd^A$ smaller than $\gd_c$ 
	will not necessarily lead to a homogeneous velocity profile $P^A$, 
	with the expected stress $\sigma^A$: 
	the flow can separate into a blocked region and a flowing region
	(profiles $P^A_2$ or $P^A_3$).  
	The local shear rate is then faster ($\gd^{\rm loc}_{A2}>\gd^A$ 
	and $\gd^{\rm loc}_{A3}>\gd^A$), 
	which corresponds to a higher stress 
	($\sigma^A_2>\sigma^A$ and $\sigma^A_3>\sigma^A$). 
	Besides, for an average shear rate  $\gd^B$ greater than $\gd_c$, 
	the flow is homogeneous again, which corresponds 
	to the expected stress $\sigma^B$ 
	(greater than $\sigma_y$).}
	\label{Fig:courbe_ecoulement_et_bandes}
	\end{figure}

\subsection{Stationary flow curve and inhomogeneous flow}

Let us now turn back to shear banding.
For such a peculiar flow to be observed,
the {\em same material} submitted to the {\em same shear stress} $\sigma_{xy}$ must be simultaneously {\em in two different deformation states}. 
As discussed for many years for various complex 
fluids~\cite{banding_phasetransition,Berret_micelles_review_molecular_gels_2006,Spenley_1993_PhysRevLett.71.939}, 
a mathematical condition for this to be possible 
is the existence of an unstable zone
in the local flow curve $\sigma_\infty(\gd)$ of the material:
it must be {\em non-monotous}.

In the case of foams, nevertheless, such an unstable portion in the flow curve itself does not exist:
how can shear bands with different shear rates coexist?

Foams and emulsions are instances of yield stress fluids, 
so that there exists a minimal value $\sigma_y$ of the stress $\sigma_{xy}$ 
below which no stationary flow occurs. 
Now when we shear the material, imposing the {\em shear rate}, 
the material {\em has} to flow, even for very small $\gd$. 
The intrinsic flow curve thus possesses 
an extrapolation in stress when $\gd\rightarrow 0$. 
Let's denote it by $\sid$. 
Note that  $\sigma_y$ and $\sid$ pertain to the {\em local} rheology curve $\sigma_\infty(\gd)$, 
not to the {\em effective, macroscopic} stationary curve 
as can be measured for example in a rheometer. 
In this discussion the flow is {\em homogeneous}. 
But the relative values of $\sid$ and $\sigma_y$, 
pertaining to the local flow curve, 
will give us hints about possible conditions for shear banding.

Let us now consider the result of a measurement 
made on a sample of this material, 
sheared in a (parallel or very low curvature)
Couette cell under imposed shear rate. 
If $\sid \geq \sigma_y$, all parts of the sample will flow, 
even at low shear rates, 
since the corresponding stress is necessarily 
everywhere greater than the yield stress. 
As mentionned before, since the flow curve 
has no intrinsic instability for higher $\gd$ values, 
no mechanism is available for shear banding.

The situation is different if $\sid < \sigma_y$. 
If we put on the same graph the yield stress $\sigma_y$ 
and the intrinsic stationary flow curve
(figure~\ref{Fig:courbe_ecoulement_et_bandes}), 
it is immediately apparent that this configuration 
allows for the coexistence of zones undergoing shear 
at rates $\gd$ such that $0 < \gd \leq \gd_c$, 
and of blocked zones remaining in the elastic regime at $\gd=0$.
The mechanism is essentially the same 
as in the classical case of instability in the flow curve 
(see figure~\ref{Fig:courbe_ecoulement_et_bandes}). 
Of course, as soon as $\gd>\gd_c$ all regions flow, 
since  $\gd>\gd_c$ implies that {\em some} regions 
flow {\em faster} than  $\gd_c$. 
The stress in these regions, as given by the flow curve, 
has to be above the yield stress $\sigma_y$. 
And since the shear stress is the same in the entire material, 
all regions support a stress greater than $\sigma_y$ 
and no region can be blocked.

\subsection{Large elastic deformations {\em versus} extra dynamic variables}
\label{Sec:large_eldef_dyn_var}

But is the situation where $\sid \neq \sigma_y$ actually possible? 
In various complex fluids, the answer is known to be yes. 
The usual explanation of such a flow curve 
is to invoque an internal extra variable 
(of a structural nature in general) 
which is coupled to the flow. 
As an example, in a simplified vision, 
this extra parameter 
can take one of two values:
flowing or non-flowing. 
Thus the stationnary curve extrapolating to $\sid$ at low $\gd$, 
and the yield stress value $\sigma_y$ 
correspond in reality to two different materials,
hence $\sid$ and $\sigma_y$ can differ. 

But as mentioned in Section~\ref{Sec:foam_rheo},
foams differ from many other complex fluids
in that the deformation that must be reached to trigger plastic flow is large.
This feature turns them into an intrinsically {\em tensorial} material.

In a stationary situation where shear banding is present,
stress conservation implies
that the shear stress $\sigma_{xy}$ is constant along the direction of the velocity gradient,
as well as the stress component $\sigma_{yy}$.
By contrast, the extra components of the stress $\sigma_{xx}(y)$ and $\sigma_{zz}(y)$ 
may vary in an arbitrary manner along the direction of the velocity gradient.
Among these, $\sigma_{xx}(y)$ is present even in a purely 2D system.
These extra components will qualitatively play the same role 
as an extra structural variable in changing the local nature of the material,
when viewed as a 1D material (along direction $y$). 

But that only explains how it is {\em possible} for shear bands to be present.
The reason why the flow curve {\em actually} extrapolates
below the yield stress at vanishing shear rates ($\sid < \sigma_y$) 
in some tensorial models, thus allowing shear banding, 
has been shown by Raufaste {\em et al.}~\cite{raufaste_2010_0732}:
as long as the material remains elastic, 
the local deformation tensor is transported by the shear flow 
along a path that is not locally aligned with itself:
the principal axes of the particulate time-derivative of the deformation
do not coincide with those of the deformation.
Hence, once plasticity is triggered, it alters the deformation evolution
until it progressively reaches the locus
where it is aligned with its transport under shear.
At least in simple examples of elasticity and plasticity,
this migration from the elastic path to the asymptotic locus
is the origin of the shear stress overshoot observed during transients~\cite{raufaste_2010_0732}.
When the plastic flow is triggered rather abruptly,
the system is still elastic just before the maximum of this overshoot,
and the asymptotic shear stress value can then lie below the last elastic shear stress value.
In other words, $\sid < \sigma_y$.

\subsection{History dependent shear bands}

Despite some similarities, 
the analogy with systems characterised by unstable flow curves 
has some limitations. 
In the case of yield stress fluids, 
there is no unstable range in $\gd$, 
which would {\em impose} phase separation 
between two phases at different flow rates. 
Shear bands are {\em possible} but not {\em necessary}. 
Also, no lever rule-like criterion 
can exist to select the relative fraction of the different bands, 
as have been argued in some fluid 
systems~\cite{Olmsted_Lu_PRE_1999.60.4397,banding_phasetransition}.

Rather, the initial distribution of $\sigma_{xx}(y)$ and $\sigma_{zz}(y)$ in the material 
will be of primary importance in the appearance of shear bands
even though the material {\em per se} remains homogeneous.
In other words, it is the {\em material history} that will lead 
to a particular flow profile. 
We will see that the initial distribution 
of stress in the material will determine the band structure.

\subsection{0D flow curve and shear banding criteria}\label{Sec:0D_banding_criteria}
\begin{figure}[ht!]
	\begin{center}
	\resizebox{0.9\columnwidth}{!}{
	\includegraphics{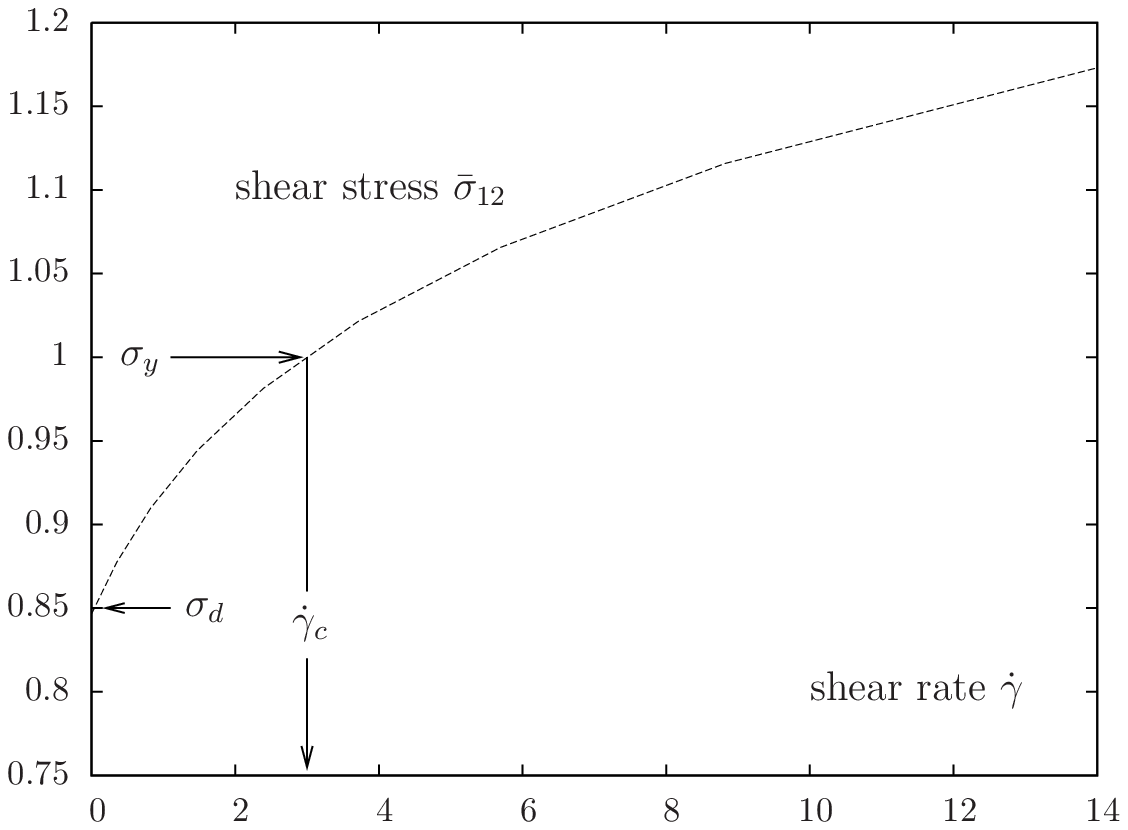}
	}
	\resizebox{0.9\columnwidth}{!}{
	\includegraphics{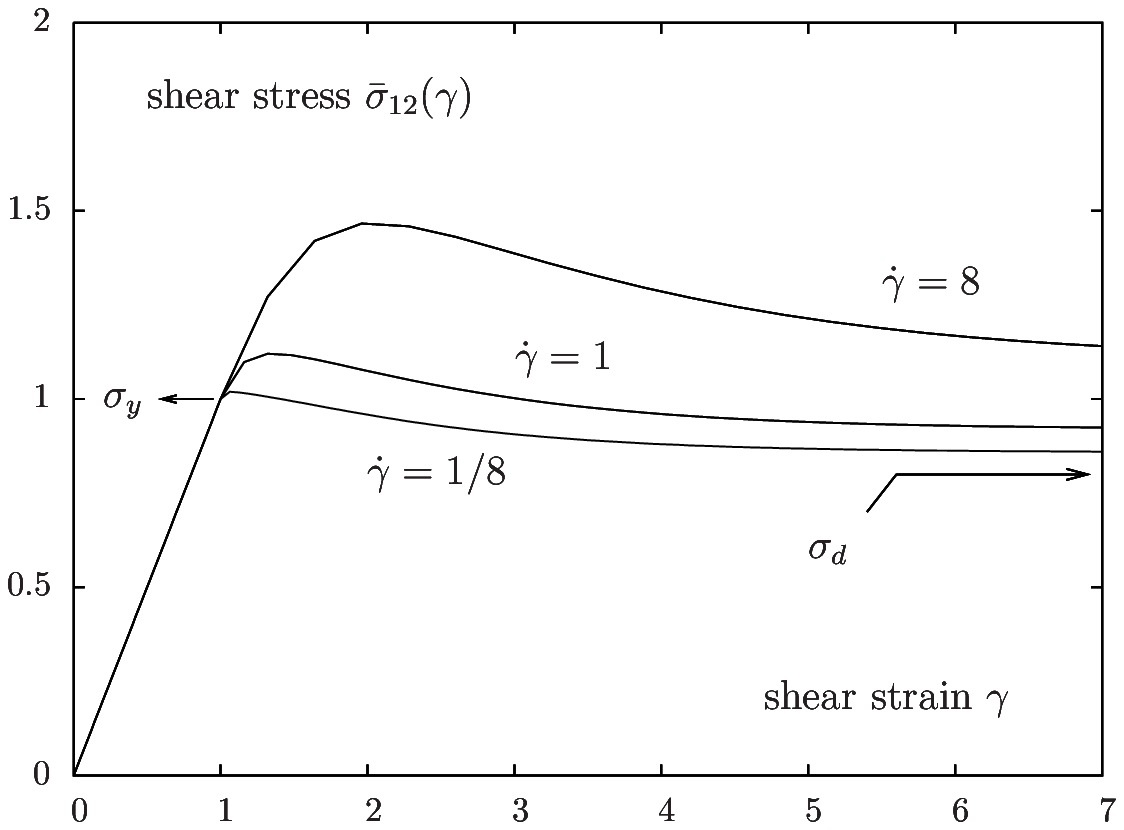}
	}
	\end{center}
	\caption{Top: typical stationary flow curve.
	Points corresponding to $\sid$,  $\sigma_y$ and $\gd_c$
	are reported on the curve. 
	Bottom: stress time evolution 
	for different imposed shear rates
	below or above $\gd_c$.}
	\label{Fig:courbe_ecoulement}
	\end{figure}

As long as the flow in the material is homogeneous, 
a local rheological model will be sufficient to describe it.
We begin by showing the typical flow curve 
corresponding to our model
(fig.~\ref{Fig:courbe_ecoulement}). 
Note that this flow curve is obtained 
under applied shear  rate conditions.

As can be observed, the conditions 
described in the introduction 
for the appearance of shear bands are fulfilled: 
stress $\sid$ is smaller than the static yield stress $\sigma_y$. 
In the shear rate range between $0$ and $\gd_c$, 
the system has the possibility 
to split the average shear rate $\gd$ 
in different proportions 
of blocked and flowing bands.\\

Thus, in the homogeneous case, 
for any values of the parameters
$\Psi$, ${\cal K}$, $a$, $b$ and $\dot{\Gamma}$,
we can use the local rheological model
to calculate the static and dynamic thresholds, $\siy$ and $\sid$, 
and the critical shear rate $\gd_c$. 
Following the line of reasoning 
developped in the introduction, 
we can then predict the range of imposed shear rates 
$[0,\,\gdc]$ inside which shear bands are {\em possible}. 

The value of $\siy$ can be obtained easily 
by simulating the system in the elastic regime
($\epspp=0$) up to the threshold ($\WW_y(\Finger)=0$), 
which corresponds to a state of the system
characterized by eigenvalues $\beta_1^y$ and $\beta_2^y$
of tensor $\Finger$,
a state for which $\siy$ can be calculated.

The different stationary state values 
of the shear stress could then obtained independently 
by continuing the simulation beyond the threshold 
in the plastic regime for each value of $\gd$, 
waiting for the stationnary value of the system  
(${\rm d}\Finger/{\rm d}t\approx 0$). 
The dynamic threshold  $\sid$ would then correspond 
to the limit of $\si_{12}$ for small $\gd$. 
The critical shear rate $\gdc$
would be obtained when the stress applied to the system 
in the stationary state would precisely correspond 
to the plastic threshold: 
$\si_{12}^{\rm stat}(\gdc)=\siy$. 
Such a procedure is natural, 
but requires successive simulations of the system 
for a large number of $\gd$ values.

\begin{figure}[htbp]
	\begin{center}
	\resizebox{1.0\columnwidth}{!}{%
	\includegraphics{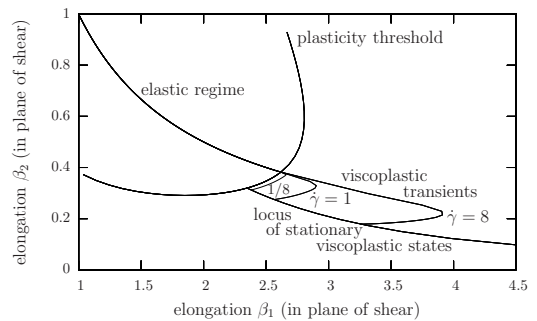}
	}
	\end{center}
	\caption{Form of the stored elastic deformation
	in the course of an experiment
	and in the stationary regime,
	for three different values
	of the shear rate $\gd$.
	The axes are the first two eigenvalues,
	$\beta_1$ and $\beta_2$, of tensor $\Finger$.}
	\label{cisaillement_beta1_beta2_k1_17_k2_67}
	\end{figure}

We have used a more direct 
approach~\cite{benito_2008_225_elasto_visco_plastic_foam}
to obtain $\sid$ and $\gdc$ (see Appendix). 
This method relies on the description 
of the evolution of the system 
in terms of independent eigenvalues 
$\beta_1$ and $\beta_2$ of tensor $\Finger$
(see figure~\ref{cisaillement_beta1_beta2_k1_17_k2_67}).

With the help of this procedure, 
the three observables which are important 
for the prediction of shear bands, 
$\siy$, $\sid$, and $\gdc$, 
are obtained directly without the need 
for simulating separately all the points 
along the stationary flow curve.

\subsection{Spatial (1D) simulations of stationary flow regimes}
\label{Sec:regime_ecoul_inhomog_stat}

As already mentioned, the discussion in the previous paragraph only provides
{\em necessary} conditions for the appearance of shear bands.
In the 1D simulations that will be discussed 
in the present section,
flow inhomogeneities will indeed sometimes emerge in a full spatial 
simulation of our tensorial model in 1D spatial dimension plus
time.

We simulate the full tensorial model in 1D using the equations of
paragraph~\ref{rheological_model}. The technical details of the
numerical scheme can be found in ~\cite{these_sylvain_benito_2009}.
{From} a numerical point of view,
let us remark in particular that 
we have checked the the grid used in the discretisation of the equations is fine enough
for all simulations presented here.

\subsubsection{Discussion of the conditions for inhomogeneous flow}

The model that we simulate
only contains material parameters 
that are {\em homogeneous} in the sample.
Hence, if the {\em initial conditions} of the flow
are also homogeneous,
the entire evolution will remain homogeneous.
Although performing a 1D simulation as a set
of partial differential equations,
we would obtain the exact same results
as in the previous Section.

In other words, since the parameters of the model
do not vary in space, shear bands can only appear
if initial conditions are, in one way or another,
inhomogeneous.

Of course, as mentioned in the Introduction,
inhomogeneities could appear in a natural way
through an extra state variable coupled to the flow,
such as the concentration.
This variable could then vary in space
and be coupled to the flow.
Concerning concentration (a conserved variable),
let us mention dilatancy phenomena,
imagined for foam~\cite{weaire_2003_2747,rioual_2005_117},
observed experimentally~\cite{marze_2005_121}
and interpreted in a geometrical 
manner~\cite{dilatancy_geometry_letter_2008,dilatancy_geometry_article}. 
Alignement (a non-conserved variable)
is another possibility. 
It has been invoked in the case
of wormlike micelle or rigid rod 
solutions~\cite{Olmsted_Lu_PRE_1999.60.4397}.

Here, we focus on inhomogeneous 
static strain/stress initial conditions,
without invoking additional variables,
and we will show that they can induce
the appearance of persistent inhomogeneities
in the flow profile.

The reason for which these initial strain inhomogeneities
can induce the appearance of blocked bands 
can be qualitatively undestood 
by considering the flow threshold $\cal{K}$.
Indeed, the stresses generated by the shear
combine with the initial stress distribution due to strain
inhomogeneities.
Depending on its orientation,
the initial stress thus precipitates or delays
the triggering of the plastic flow.

\subsubsection{Initial inhomogeneous strain distribution}

First, the existence of stress inhomogeneities
stored in the system before it is set into motion
is physically well motivated.
For instance, introducing a foam sample
into an apparatus requires non-homogeneous flows.
Inhomogeneous stresses will build up in the sample
very likely, except if particular care is taken,
such as a slow, {\em in situ} drying 
of an initially wet foam.

We will always assume that the initial state is at rest, 
that is, that the elastic stresses 
are at equilibrium in the sample.
However, even when this equilibrium is imposed,
there exist a large set of possible initial 
spatial distributions of stresses and strains.
For example, if the system is invariant
in the $xz$ plane of the shearing walls,
some components of the stress must be homogeneous.
That is the case for $\sigma_{xy}$, 
$\sigma_{yy}$ and $\sigma_{yz}$. 
The other stress components, however,
can freely vary as a function of $y$
as long as they remain constant in each $xz$ plane.
It thus corresponds to a 1D inhomogeneity
in the direction of the velocity gradient.

In this paragraph, we examine
a very simple case of initial condition,
with uniaxial extension along axis $x$, with $\Finger_{xx}=\Finger_{xx}(y)$
and $\Finger_{yy}=\Finger_{zz}=1/\sqrt{\Finger_{xx}}$.
We used a simple monotonic function:
\be
\label{eq:Bxx_beta_croissante}
\Finger_{xx}=1.1+\epsilon \,y^\beta\,(1-(1-y)^\beta).
\ee

In practice, in order to prepare a sample in such a state,
one must compress the foam in a non-homogeneous manner.
Typically, a block of foam with a trapezoidal shape
forced to take a rectangular shape will undergo this kind of strain inhomogeneity.
In this context, we cannot comment on any relation between these strain inhomogeneities in our 
{\em continuum} model and the {\em local structural disorder} existing at the bubble level,
as this disorder {\em is averaged out} in our continuum description.
In particular, there is no clear structural interpretation of the amplitude $\epsilon$ 
of the strain inhomogeneities in the prepared sample.

\subsubsection{Characterizing the inhomogeneous flows}

A typical sequence of velocity profiles
obtained in our numerical simulations
displays as follows.
The velocity profile is initially homogeneous.
It remains homogeneous as long as the entire sample
is in the elastic regime.
The regions where the initial stress is the highest
in the direction of the applied deformation
reach the threshold first.
The average shear rate being constant,
this onset of creep leads both to 
a higher shear rate in the creeping regions
and to a lower one in the others.
The high shear rate then induces
the saturation of the stress due to creep,
and the shear becomes blocked
in the region below the threshold.
In the stationary regime,
a blocked band coexists with a sheared band
at the same shear stress.

In the corresponding transient regime,
non-trivial phenomena may appear,
especially at the boundary of the blocked zone.
Transient negative local shear rates are observed 
due to stored elastic stresses.

Let us now address the characteristics
of the stationary velocity profile,
again from the behaviour of the local rheological model.

The first feature of interest
is that in the flowing regions,
the velocity profile is linear,
that is, the shear rate is uniform.
That can be understood in the following manner.
All the regions which, in the stationary state,
respond through a non-zero shear rate,
correspond to a point located on the stationary flow curve
in the $\beta_1$-$\beta_2$ diagram 
of figure~\ref{cisaillement_beta1_beta2_k1_17_k2_67}.
Each point of this curve corresponds 
to a different shear stress.
Thus, since each layer of the flow
undergoes the same shear stress,
they all actually correspond 
to the same point on the curve
and thus respond through the same shear rate.

A second feature results from the fact that in space,
while $\sigma_{xy}$ and $\sigma_{yy}$ are continuous,
$\sigma_{xx}$ and $\sigma_{zz}$ 
can perfectly be discontinuous.
That is precisely the case at the boundary
between a shear and a blocked region.
This is the flow counterpart
of the discontinuity in the $\beta_1$-$\beta_2$ diagram,
between the points below the threshold
and the point with a stationary shear rate
that corresponds to the flowing region.
Actually, the only coupling between the different layers
comes from the fact that 
{\em (i)} $\sigma_{xy}$ and $\sigma_{yy}$ 
must be every where the same,
and {\em (ii)} the integral of $\gd$ 
over the gap thickness is fixed 
by the imposed wall velocity.
As a consequence, in an inhomogeneous flow,
the organisation of the blocked and flowing layers
is not unique:
any permutation of the layers is actually possible.
Again, initial conditions decide
upon the particular structure adopted by the flow.
Two initial conditions corresponding 
to permutated layers would lead
to the same permutation in the stationary flow structure.

\subsection{Parameters affecting the existence of blocked bands}

In this section, we want to describe, 
within the parameter space  ($\Psi$, ${\cal K}$, $a$, $b$, $\dot{\Gamma}$) 
the regions inside which shear bands are possible.  
These domains will be represented through sections 
in five different planes: $(\cal K$,$\dot{\Gamma})$, $(\Psi$,$\dot{\Gamma})$, $(a,b)$ and $(\Psi$, $\cal K)$.
The results are presented 
in figures~\ref{Fig:Banding_Kappa_Psi},
\ref{fig:Gdot_Kappa}, \ref{fig:Psi_Gdot},
\ref{fig:Kappa_a}, and \ref{fig:a_b}.

As will be discussed below (paragraph~\ref{sec:dep_init_cond}),
the choice of the initial conditions can have a crucial impact
on the existence of shear bands.
A complete investigation of the model
would therefore require a very thorough exploration
not only of the parameters ($\Psi$, ${\cal K}$, $a$, $b$, $\dot{\Gamma}$) 
but also of the shape and amplitude of the initial strain profile.
In order to favour the appearance of shear bands
without needing to refine the exploration of various strain profiles,
we selected a very large amplitude $\epsilon=500\%$
for the shape mentioned in Eq.~(\ref{eq:Bxx_beta_croissante}).
This applies to Figs~\ref{Fig:Banding_Kappa_Psi}-\ref{fig:a_b}.

\subsubsection{$(\cal K$, $\Psi)$ plane}

\begin{figure}[htbp]
	\begin{center}
	\begin{psfrags}
	\psfrag{Kappa}{${\cal K}$}
	\psfrag{Psi}{${\Psi}$}
	\psfrag{GammaDot=0.05   a=b=1/7}{$\dot{\Gamma}=0.05$, $a=b=1/7$}
	\includegraphics[width=8cm]{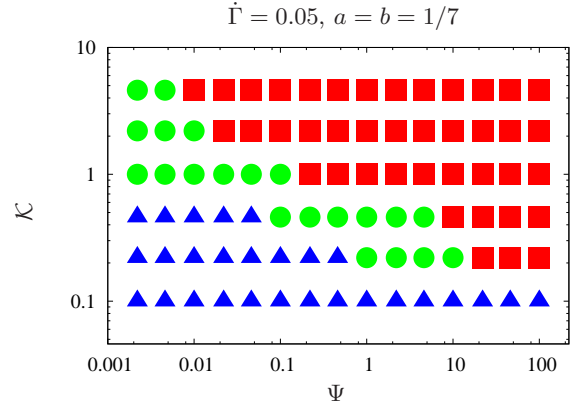}
	\end{psfrags}
	\end{center}
	\caption{Comparison of 0D and 1D simulations
	in plane ($\cal K$, $\Psi$), using $a=b=1/7$ and $\dot\Gamma=0.05$.
	Blue triangles indicate values for which the 0D model
	allows only uniform flow.
	Red squares indicate values for which shear banding
	was obtained in the 1D simulation for the initial conditions chosen.
	Green disks indicate additional values 
	for which the 0D model allows shear banding.}
	\label{Fig:Banding_Kappa_Psi}
	\end{figure}

In the $(\cal K$, $\Psi)$ plane, 
bands are predicted by the local model for large values of $\Psi$ and $\cal K$.
Note that in the figure, zones where no bands can appear are denoted by blue triangles.
Small values of $\Psi$ correspond to a situation where the relaxation time $\frac{\eta_s }{G}$ 
in the absence of plasticity is far smaller than the relaxation time $\tau$ corresponding to the plasticity. 
It is thus a regime dominated by
the fluid viscosity, where the creep plays no role. As $\Psi$ increases, 
the creep becomes dominant, and shear bands can appear for lower values of
the threshold ($\cal K$) (figure~\ref{Fig:Banding_Kappa_Psi}).

As expected, the shear bands observed in the simulations appear only in regions authorized 
by the scalar model. 

The green dots correspond to values for wich the scalar model allows the presence of shear bands, wich are not observed in the 
simulations for a specific set of initial conditions. As will be commented further on, the extent of this green zone depends on these initial conditions, demonstrating one of 
the main points of this work.

\subsubsection{$(\cal K$, $\dot{\Gamma})$ plane}

\begin{figure}[htbp]
	\begin{center}
	\begin{psfrags}
	\psfrag{Kappa}{${\cal K}$}
	\psfrag{GammaDot}{$\dot\Gamma$}
	\psfrag{Psi=0.1   a=b=1/7}{$\Psi=0.1$, $a=b=1/7$}
	\includegraphics[width=8cm]{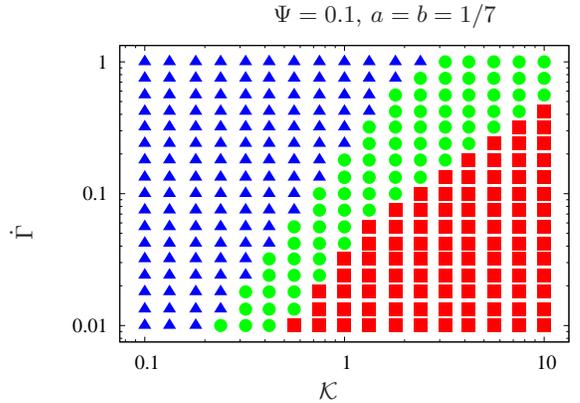}
	\end{psfrags}
	\end{center}
	\caption{Comparison of 0D and 1D simulations
	in plane ($\dot\Gamma$, $\cal K$), using $a=b=1/7$ and $\Psi=0.1$.
	Blue triangles indicate values for which the 0D model
	allows only uniform flow.
	Red squares indicate values for which shear banding
	was obtained in the 1D simulation for the initial conditions chosen.
	Green disks indicate additional values
	for which the 0D model allows shear banding.}
	\label{fig:Gdot_Kappa}
	\end{figure}

In the $(\cal K$, $\dot{\Gamma})$ plane, 
bands should be predicted for small values of $\dot{\Gamma}$ 
(due to the small velocities wich explore regions of the flow curve close to the origin in $\dot{\Gamma}$), 
and for large values of $\cal K$.
Indeed, in that case the static threshold is large
which favours bands since they are possible 
{\em below} this threshold (figure~\ref{fig:Gdot_Kappa}).

Concerning the relation between the values predicted for shear bands in the 0D model
and the observations in the 1D simulations, the same remarks hold as for the previous paragraph.

\subsubsection{$(\Psi$,$\dot{\Gamma})$ plane}

\begin{figure}[htbp]
	\begin{center}
	\begin{psfrags}
	\psfrag{Kappa}{${\cal K}$}
	\psfrag{GammaDot}{$\dot\Gamma$}
	\psfrag{Psi}{${\Psi}$}
	\psfrag{K=1   a=b=1/7}{${\cal K}=1$, $a=b=1/7$}
	\includegraphics[width=8cm]{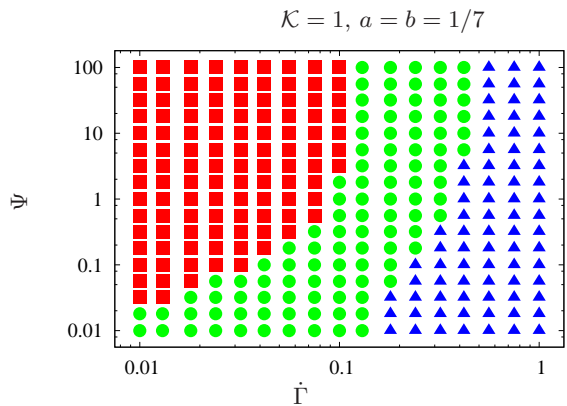}
	\end{psfrags}
	\end{center}
	\caption{Comparison of 0D and 1D simulations
	in plane ($\dot\Gamma$, $\Psi$), using $a=b=1/7$ and ${\cal K}=1.0$.
	Blue triangles indicate values for which the 0D model
	allows only uniform flow.
	Red squares indicate values for which shear banding
	was obtained in the 1D simulation for the initial conditions chosen.
	Green disks indicate additional values 
	for which the 0D model allows shear banding.}
	\label{fig:Psi_Gdot}
	\end{figure}

Observations in this plane corroborate the analysis in the two previous planes : 
bands are allowed (and are observed) for low $\dot{\Gamma}$ values and high
$\Psi$ values (figure~\ref{fig:Psi_Gdot}). 

\subsubsection{$(\cal K$, $a)$ plane}

Again bands appear for large values of $\cal K$. The influence of the 
$a$ parameter is far more subtle to assess, being related to non-trivial tensorial effects
of the elastic ($a$) and plastic ($b$ taken as $1-a$ here) terms.

The same remarks hold concerning the correlation between the 0D model and 1D simulations.

	\begin{figure}[htbp]
	\begin{center}
	\begin{psfrags}
	\psfrag{a = 1 - b}{$a=1-b$}
	\psfrag{K}{${\cal K}$}
	\psfrag{Psi=0.1   GammaDot=0.05   b=1-a}{$\Psi=0.1$, $\dot\Gamma=0.05$, $b=1-a$}
	\includegraphics[width=8cm]{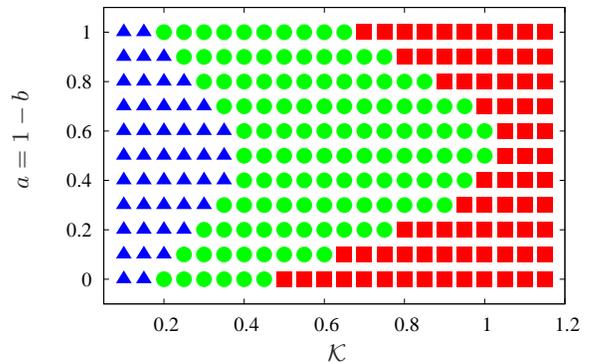}
	\end{psfrags}
	\end{center}
	\caption{Comparison of 0D and 1D simulations
	in plane ($\cal K$, $a$), using $b=1-a$, $\dot\Gamma=0.05$ and $\Psi=0.1$.
	Blue triangles indicate values for which the 0D model
	allows only uniform flow.
	Red squares indicate values for which shear banding
	was obtained in the 1D simulation for the initial conditions chosen.
	Green disks indicate additional values 
	for which the 0D model allows shear banding.}
	\label{fig:Kappa_a}
	\end{figure}

\subsubsection{$(a,b)$  plane}

\begin{figure}[htbp]
	\begin{center}
	\begin{psfrags}
	\psfrag{a}{$a$}
	\psfrag{b}{$b$}
	\psfrag{Psi}{${\Psi}$}
	\psfrag{Psi=0.1   GammaDot=0.05   K=0.3}{$\Psi=0.1$, $\dot\Gamma=0.05$, ${\cal K}=0.3$}
	\includegraphics[width=8cm]{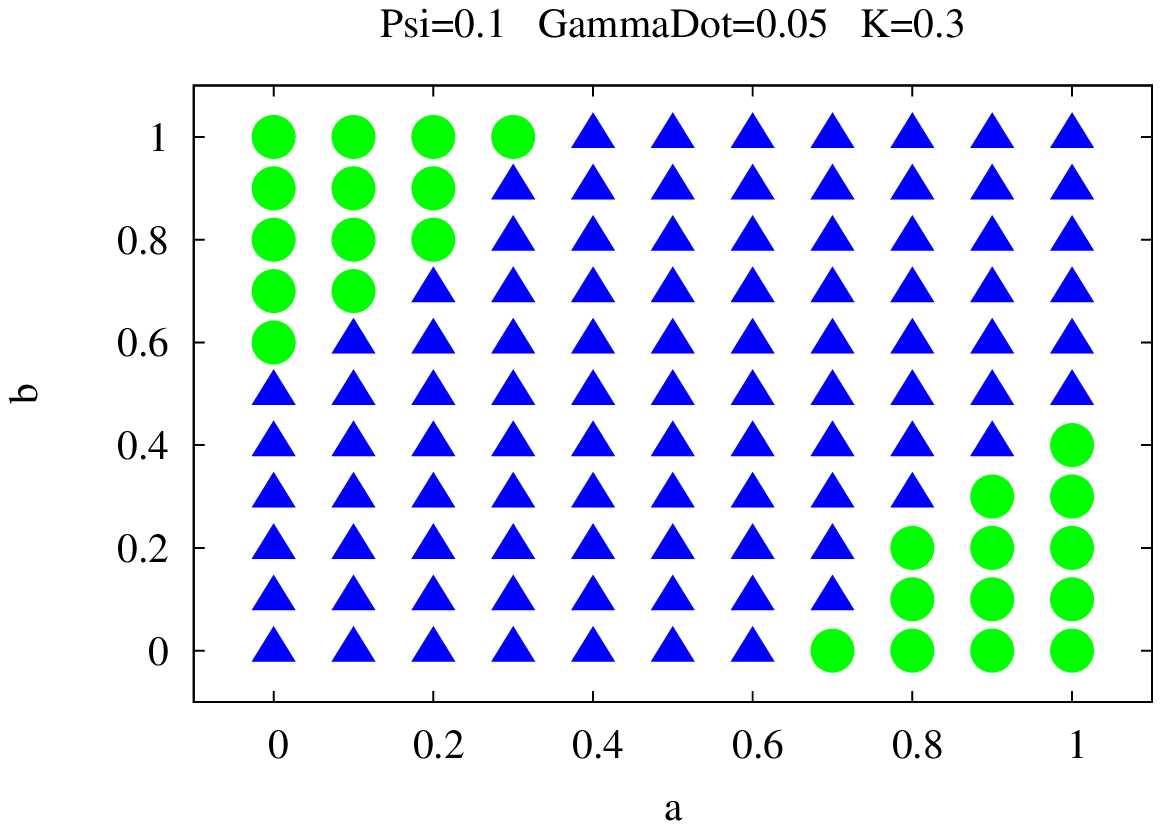}
	\end{psfrags}
	\end{center}
	\begin{center}
	\begin{psfrags}
	\psfrag{a}{$a$}
	\psfrag{b}{$b$}
	\psfrag{Psi}{${\Psi}$}
	\psfrag{Psi=0.1   GammaDot=0.05   K=0.8}{$\Psi=0.1$, $\dot\Gamma=0.05$, ${\cal K}=0.8$}
	\includegraphics[width=8cm]{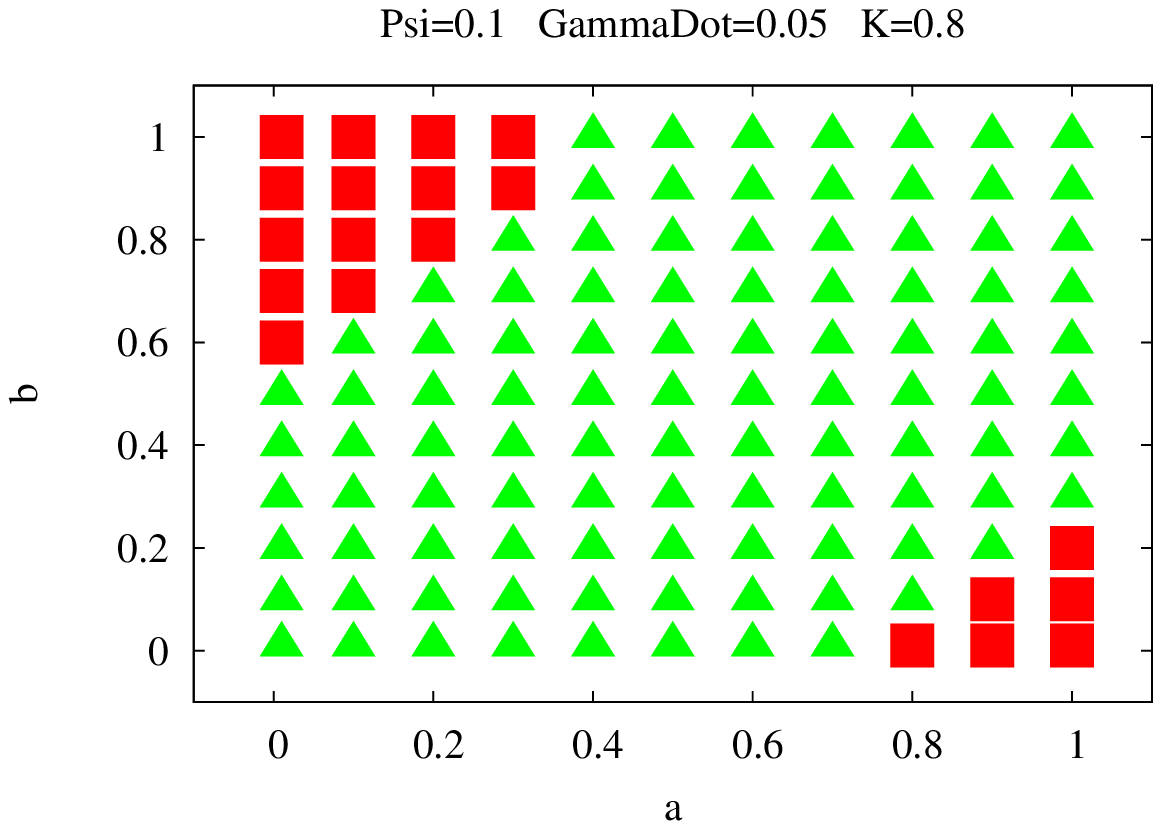}
	\end{psfrags}
	\end{center}
	\caption{Comparison of 0D (top) and 1D (bottom) simulations
	in plane ($a$, $b$), using $\dot\Gamma=0.05$ and $\Psi=0.1$.
	Following the indications of Fig.~\ref{fig:Kappa_a}
	we chose ${\cal K}=0.3$ for 0D simulations
	and ${\cal K}=0.8$ for 1D simulations.
	The second diagonals ($b=1-a$) in the present diagrams
	correspond to vertical lines in Fig.~\ref{fig:Kappa_a}.
	Blue triangles indicate values for which the 0D model
	allows only uniform flow.
	Red squares indicate values for which shear banding
	was obtained in the 1D simulation for the initial conditions chosen.
	Green disks indicate additional values 
	for which the 0D model allows shear banding.}
	\label{fig:a_b}
	\end{figure}

Finally, in the ($a$, $b$) plane, 
one is again confronted with 3D effects 
which are difficult to discuss in intuitive terms 
(figure~\ref{fig:a_b}). 
The way the elasticity (parameter $a$) 
and the plastic deformation rate (parameter $b$) 
are coupled in a tensorial way affects
the critical rate $\gdc$ 
and can be enough to eliminate all possibilities 
of shear bands.

\subsubsection{Dependance on the initial conditions}
\label{sec:dep_init_cond}

We have always observed that the regions in which blocked bands
actually appeared in the 1D simulations
are strictly included, as expected, 
in the regions authorized by the local rheological model.

However, the boundary of these regions do not coincide.
In fact, for the same values of the parameters,
the extent of the banding zone depends crucially on the initial conditions, while always remaining in the 
region allowed by the rheological model.
In other words, the behaviour of the system is history dependent,
a feature realized independently in a recent work
on a related tensorial model with plasticity~\cite{cheddadi_JRheol2012_213}.

\begin{figure}[htbp]
	\begin{center}
	\begin{psfrags}
	\psfrag{a}{$a$}
	\psfrag{b}{$b$}
	\psfrag{Psi}{${\Psi}$}
	\psfrag{Psi=0.1   a=b=1/7}{$\Psi=0.1$, $a=b=1/7$}
	\psfrag{Kappa}{${\cal K}$}
	\psfrag{GammaDot}{$\dot\Gamma$}
	\includegraphics[width=1.0\columnwidth]{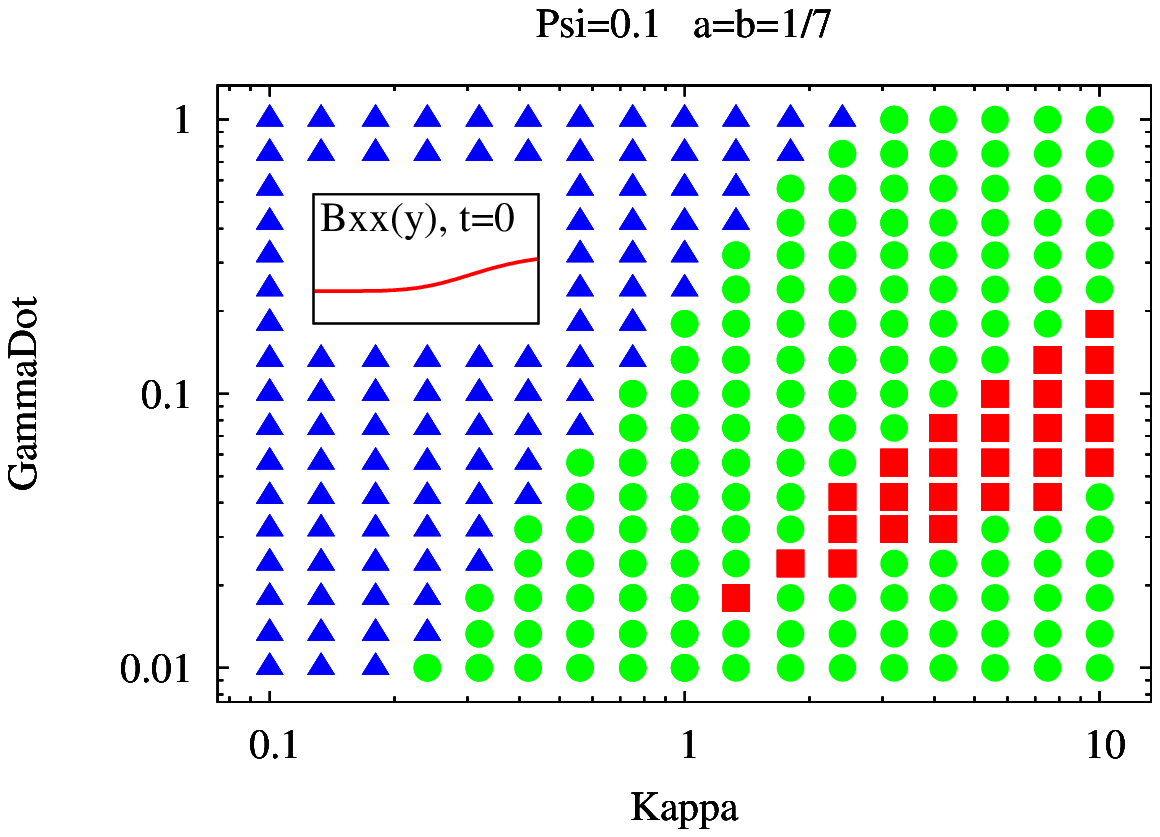}
	\end{psfrags}
	\end{center}
	\begin{center}
	\begin{psfrags}
	\psfrag{Psi}{${\Psi}$}
	\psfrag{a}{$a$}
	\psfrag{b}{$b$}
	\psfrag{Kappa}{${\cal K}$}
	\psfrag{GammaDot}{$\dot\Gamma$}
	\psfrag{Psi=0.1   a=b=1/7}{$\Psi=0.1$, $a=b=1/7$}
	\includegraphics[width=1.0\columnwidth]{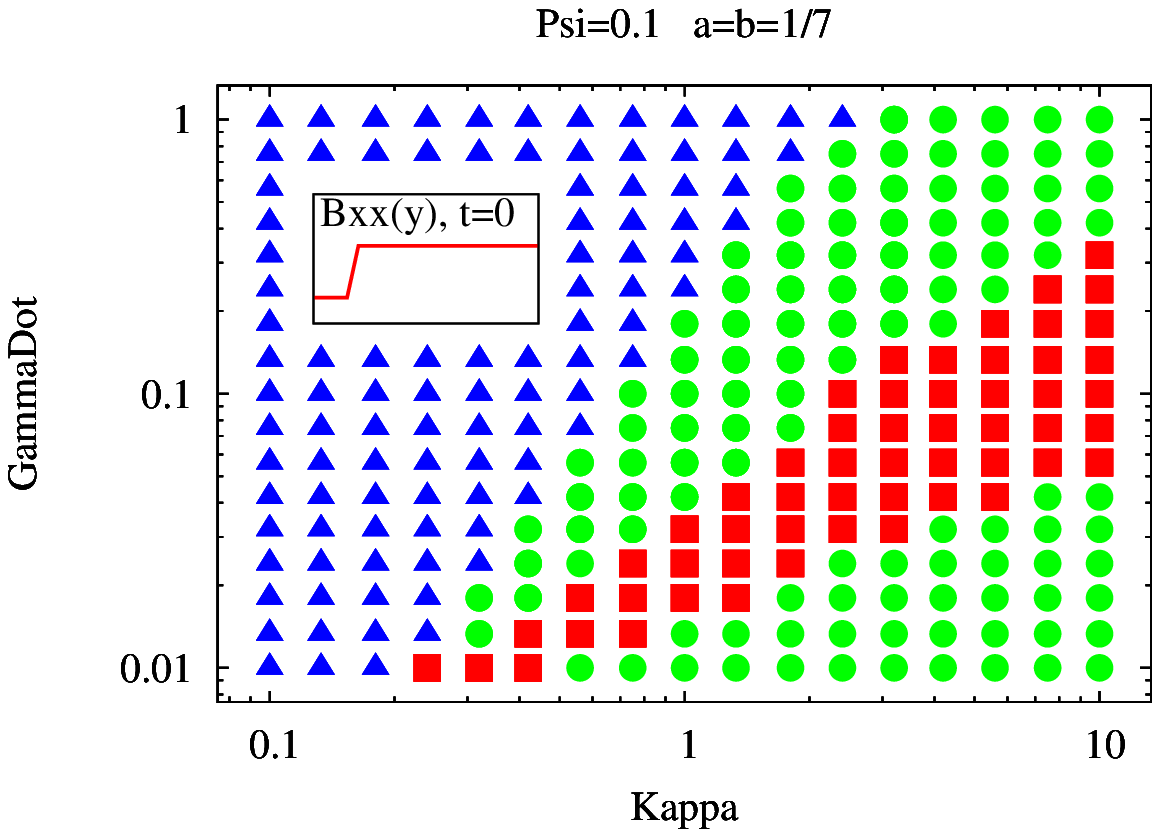}
	\end{psfrags}
	\end{center}
	\caption{Initial conditions dependency in the plane ($\dot\Gamma$, $\cal K$), using $a=b=1/7$ and $\Psi=0.1$. 
In the upper graph we considered sigmoidal initial conditions, and in the lower graph step-like ones. In both cases, the amplitude 
of the strain inhomogeneities was reduced to 10\% in amplitude, whereas Fig.~\ref{fig:Gdot_Kappa} corresponded to 500\% to enhance the effect.}
	\label{fig:in_cond_dependency}
	\end{figure}

To illustrate that, we have varied both the form and the amplitude
of the spatial modulation of the initial deformation.
In Figures~\ref{Fig:Banding_Kappa_Psi}-\ref{fig:a_b},
the initial profile was given by the non-linear form
of Eq.~(\ref{eq:Bxx_beta_croissante}) with $\epsilon=500\%$.
By contrast, in Figure~\ref{fig:in_cond_dependency}, 
for the upper graph we chose a simple sigmoidal profile
centered around a selected altitude $y_0$,
\be
\Finger_{xx}=1+\epsilon \,y^{\beta}\,\frac{y_0^\beta+1}{y_0^\beta+y^\beta},
\ee
and for the lower graph we chose a step-like function,
both with $\epsilon=10\%$.
Comparing figures~\ref{fig:Gdot_Kappa} and~\ref{fig:in_cond_dependency}
shows that the parameter domain where shear bands actually appear
can depend in a non-trivial manner
not only on the shape but also on the amplitude
of the initial strain profile.

Actually, we expect that a thorough exploration
of the region authorized for the shear bands
could be achieved through a very fine adjustment
of the initial condition profile for each set of parameters.


\section{Conclusion}

The present study on shear bands in liquid foams
was conducted on a rheological model~\cite{benito_2008_225_elasto_visco_plastic_foam}
whose predictions we here compare
to existing rheological measurements under large amplitude oscillations
(see Section~\ref{sec:comparison_with_experiments}).

The shear bands obtained with the model
in parallel Couette geometry
display several somewhat unusual features.
\begin{enumerate}
\item The shear bands depend on the initial conditions.
More generally, the stationary state is history dependent:
only the flowing regions coincide with the simple linear velocity profile
obtained in the case of a stationary homogeneous flow.
\item The response of the model in a stationary homogeneous flow
is continuous when approaching zero shear rate,
with no forbidden region below some finite shear rate.
\item The model does not contain any non-conserved (structural) order parameter.
\end{enumerate}

This study thus shows that shear bands 
can arise naturally in a fully tensorial rheological model.
This departs from most works in the shear banding community
which put less emphasis on the tensorial character of the various models.
Here, the appearance and persistence of the bands
result from the combination of the initial conditions
and the difference between the static
and the dynamic flow thresholds (in shear geometry),
which itself arises from the tensorial character of the model.
The shear rate is discontinuous at the boundary
between the flowing and blocked regions,
but the value of the shear rate near the boundary
as well as the band widths depend on the sample history and preparation.

\section*{Acknowledgements}
We warmly thank Florence Rouyer and her co-authors
for transmitting the raw data from the large amplitude
oscillatory experiments of Ref.~\cite{Rouyer08}
discussed in Section~\ref{sec:comparison_with_experiments}.

\appendix
\section{Direct method for obtaining 
the stationary state in the local rheological model}

Let us start from the point ($\beta_1^y$, $\beta_2^y$)
and follow the plasticity threshold $\WW_y(\Finger)=0$ 
until the stationarity condition is fulfilled.
This condition can be expressed
using the following observation: 
in the stationary regime, there is no plastic flow
in the vorticity direction~\cite{benito_2008_225_elasto_visco_plastic_foam}.
In other words, the third eigenvalue 
of tensor ${\cal G}(\Finger)$ is zero:
\bee
g_3(\beta_1, \beta_2)={\cal G}_3(\beta_1, \beta_2, \beta_3)=0,
\eee
with $\beta_3=\frac{1}{\beta_1\beta_2}$.
We thus directly obtain 
the dynamic threshold ($\beta_1^d$, $\beta_2^d$)
of the system.
We then follow the same stationarity condition
$g_3(\beta_1, \beta_2)=0$ 
until we reach the desired shear stress $\si_{12}=\siy$.
We thus directly obtain the stationary state
($\beta_1^{cc}$, $\beta_2^{cc}$)
that corresponds to the critical shear rate $\gdc$. 
In practice, we follow the threshold curve using
$\hat{\WW}_y(\beta_1, \beta_2)
=\WW_y(\beta_1, \beta_2, \beta_3)=0$
(with $\beta_3=\frac{1}{\beta_1\beta_2}$)
by integrating the following differential system:
\bee
\varepsilon_{\WW_y}\,\frac{{\rm d}\beta_1}{{\rm d}t}
&=&\frac{\partial\hat{\WW}_y}{\partial\beta_2}
=\frac{\partial {\WW_y}}{\partial\beta_2}
-\frac{1}{\beta_1\beta_2^2}
\frac{\partial {\WW_y}}{\partial\beta_3}\\
-\varepsilon_{\WW_y}\,\frac{{\rm d}\beta_2}{{\rm d}t}
&=&\frac{\partial\hat{\WW}_y}{\partial\beta_1}
=\frac{\partial \WW_y}{\partial\beta_1}
-\frac{1}{\beta_1^2\beta_2}
\frac{\partial \WW_y}{\partial\beta_3}
\eee
where the sign of $\varepsilon_{\WW_y}=\pm 1$
is chosen in such a way as to follow the curve
${\WW_y}$ in the desired direction.
Similarly, we follow the curve of stationary states,
$g_3(\beta_1, \beta_2)={\cal G}_3(\beta_1, \beta_2, \beta_3)=0$
by integrating the following differential system:
\bee
\varepsilon_{g_3}\,\frac{{\rm d}\beta_1}{{\rm d}t}
&=&\frac{\partial g_3}{\partial\beta_2}
=\frac{\partial{\cal G}_3}{\partial\beta_2}
-\frac{1}{\beta_1\beta_2^2}
\frac{\partial{\cal G}_3}{\partial\beta_3}\\
-\varepsilon_{g_3}\,\frac{{\rm d}\beta_2}{{\rm d}t}
&=&\frac{\partial g_3}{\partial\beta_1}
=\frac{\partial{\cal G}_3}{\partial\beta_1}
-\frac{1}{\beta_1^2\beta_2}
\frac{\partial{\cal G}_3}{\partial\beta_3}
\eee
where the sign of  $\varepsilon_{g_3}=\pm 1$
is chosen in such a way as to follow the curve
$g_3=0$ in the desired direction.

\bibliography{shear-bands}

\end{document}